\documentclass[aps,twocolumn,amsmath,nofootinbib]{revtex4}
\usepackage{graphicx}
\usepackage{multirow}
\usepackage{dcolumn}
\usepackage{amssymb,amscd,xypic,bm,dsfont,wasysym,amsmath,mathrsfs,latexsym,psfrag,accents,mathtools}
\usepackage{float}
\usepackage{color}
\usepackage{xcolor}

\newcolumntype{L}[1]{>{\raggedright\arraybackslash}p{#1}}
\newcolumntype{C}[1]{>{\centering\arraybackslash}p{#1}}
\newcolumntype{R}[1]{>{\raggedleft\arraybackslash}p{#1}}

\RequirePackage[normalem]{ulem} 
\RequirePackage{color}\definecolor{RED}{rgb}{1,0,0}\definecolor{BLUE}{rgb}{0,0,1} 

\newcommand\tilx{\tilde{X}}
\newcommand\tilz{\tilde{Z}}
\newcommand\tilg{\tilde{\Gamma}}
\newcommand\tilu{\tilde{U}}

\newcommand\pper{P_{\sma{\perp}}}

\newcommand\myspace{\;\;\;\;}

\newcommand\Wline[2]{ \pdg{\W}_{\sma{#1 \leftarrow #2}}}

\newcommand{\bb}{\boldsymbol{b}}
\newcommand{\bk}{\boldsymbol{k}}
\newcommand{\bkone}{\boldsymbol{k_1}}
\newcommand{\bktwo}{\boldsymbol{k_2}}
\newcommand{\bp}{\boldsymbol{p}}
\newcommand{\br}{\boldsymbol{r}}
\newcommand{\bG}{\boldsymbol{G}}
\newcommand{\bR}{\boldsymbol{R}}
\newcommand{\bdelta}{\boldsymbol{\delta}}

\newcommand{\pgdel}{\pdg{g}_{\boldsymbol{\delta}}}
\newcommand{\bmz}{\bar{M}_z}

\newcommand{\calgdel}{\pdg{\breve{g}}_{\sma{\bdelta}}}
\newcommand{\brevegdel}{\pdg{\breve{g}}_{\sma{\bdelta}}}

\newcommand{\bmx}{\bar{M}_x}

\newcommand{\utg}{U_{\sma{Tg \bdelta}}}
\newcommand{\ut}{U_{\sma{T}}}

\newcommand{\ubmx}{U_{\sma{\bmx}}}

\newcommand{\caltgdel}{\breve{T}_{\sma{g \bdelta}}}

\newcommand{\hatgdel}{\pdg{\hat{g}}_{\sma{\bdelta}}}
\newcommand{\hatTg}{\pdg{\hat{T}}_{\sma{g\bdelta}}}
\newcommand\csixdel{C_{\sma{6,\bdelta}}}
\newcommand\brevecsixdel{\breve{C}_{\sma{6,\bdelta}}}
\newcommand\cthreedel{C_{\sma{3,\bdelta}}}
\newcommand\brevecthreedel{\breve{C}_{\sma{3,\bdelta}}}

\newcommand\brevecthreetwodel{\breve{C}_{\sma{3,2\bdelta}}}
\newcommand\brevectwothreedel{\breve{C}_{\sma{2,3\bdelta}}}
\newcommand\ctwothreedel{C_{\sma{2,3\bdelta}}}

\newcommand\cndel{C_{\sma{n,\bdelta}}}
\newcommand\cbarndel{C_{\sma{\bar{n},\bdelta}}}
\newcommand\ucndel{U_{\sma{C_n,\bdelta}}}
\newcommand\hatcndel{\hat{C}_{\sma{n,\bdelta}}}
\newcommand\brevecndel{\breve{C}_{\sma{n,\bdelta}}}
\newcommand\brevectwotwodel{\breve{C}_{\sma{2,2\bdelta}}}
\newcommand\brevecfourdel{\breve{C}_{\sma{4,\bdelta}}}
\newcommand\brevectwodel{\breve{C}_{\sma{2,\bdelta}}}

\newcommand{\sx}{\sigma_{\sma{1}}}
\newcommand{\sy}{\sigma_{\sma{2}}}
\newcommand{\sz}{\sigma_{\sma{3}}}

\newcommand{\tx}{\tau_{\sma{1}}}

\newcommand{\tz}{\tau_{\sma{3}}}

\newcommand{\ins}[1]{\;\;\;\;\text{#1}\;\;\;\;}

\newcommand{\kpar}{\boldsymbol{k}_{{\shortparallel}}}

\newcommand{\ang}{\mathrm{\AA}}

\newcommand{\partdif}[2]{\frac{ \partial #1}{\partial #2}}

\newcommand{\cala}{{\cal A}}

\newcommand{\calc}{{\cal C}}
\newcommand{\cald}{{\cal D}}

\newcommand{\calf}{{\cal F}}

\newcommand{\calh}{{\cal H}}
\newcommand{\cali}{{\cal I}}

\newcommand{\calk}{{\cal K}}

\newcommand{\calm}{{\cal M}}

\newcommand{\calq}{{\cal Q}}
\newcommand{\calr}{{\cal R}}

\newcommand{\calt}{{\cal T}}
\newcommand{\calu}{{\cal U}}

\newcommand{\caly}{{\cal Y}}

\newcommand{\noi}[1]{\noindent (#1)}
\newcommand{\imp}{\;\;\Rightarrow\;\;}
\newcommand{\mo}{\text{-}1}

\newcommand{\ketbra}[2]{\big|  #1  \big\rangle \big\langle #2 \big| }

\newcommand{\bra}[1]{\big\langle#1\big|}
\newcommand{\ket}[1]{\big|#1\big\rangle}

\newcommand{\bea}{\begin{eqnarray}}
\newcommand{\enea}{\end{eqnarray}}
\newcommand{\beq}{\begin{equation}}
\newcommand{\eneq}{\end{equation}}

\newcommand{\pdg}[1]{{#1}^{\phantom{\dagger}}}

\newcommand{\eq}{=&\;}
\newcommand{\ab}{\alpha\beta}

\newcommand{\low}{L$\ddot{\text{o}}$wdin$\;$}

\newcommand{\C}{{\cal C}}

\newcommand{\W}{{\cal W}}

\newcommand{\bpm}{\begin{pmatrix}}
\newcommand{\epm}{\end{pmatrix}}

\newcommand{\bal}{\begin{align}}
\newcommand{\eal}{\end{align}}

\newcommand{\R}{\mathbb{R}}

\newcommand{\dg}[1]{#1^{\scriptstyle{\dagger}}}

\newcommand{\sma}[1]{\scriptscriptstyle{#1}}

\newcommand{\noc}{n_{\sma{{occ}}}}

\newcommand{\Z}{\mathbb{Z}}

\newcommand{\qed}{\nobreak \ifvmode \relax \else
      \ifdim\lastskip<1.5em \hskip-\lastskip
      \hskip1.5em plus0em minus0.5em \fi \nobreak
      \vrule height0.75em width0.5em depth0.25em\fi}

\begin{document}

\title{Hourglass Fermions}
\author{Zhijun Wang$^{1}$,$^*$ A. Alexandradinata$^{1,2}$} 
\thanks{These authors contributed equally to this work.}
\author{R.~J. Cava$^3$}
\author{B. Andrei Bernevig$^1$}
\email{ Bernevig@princeton.edu}
\affiliation{${^1}$Department of Physics, Princeton University, Princeton, NJ 08544, USA}
\affiliation{${^2}$Department of Physics, Yale University, New Haven, CT 06520, USA}
\affiliation{${^3}$Department of Chemistry, Princeton University, Princeton, NJ 08540, USA}


\maketitle

\textbf{
Spatial symmetries in crystals are distinguished by whether they preserve the spatial origin. We show how this basic geometric property gives rise to a new topology in band insulators. We study spatial symmetries that translate the origin by a fraction of the lattice period, and find that these nonsymmorphic symmetries protect a novel surface fermion whose dispersion is shaped like an hourglass; surface bands connect one  hourglass to the next in an unbreakable zigzag pattern. These exotic fermions are materialized in the large-gap insulators: KHg$X$ ($X{=}$As,Sb,Bi), which we propose as the first material class whose topology relies on nonsymmorphic symmetries. Beside the hourglass fermion, another surface of KHg$X$ manifests a 3D generalization of the quantum spin Hall effect, which has only been observed in 2D crystals. To describe the bulk topology of nonsymmorphic crystals, we propose a non-Abelian generalization of the geometric theory of polarization. Our nontrivial topology originates from an inversion of the rotational quantum numbers, which we propose as a fruitful criterion in the search for topological materials. 
}

Spatial symmetries are ubiquitous in crystals. A basic geometric property that distinguishes these symmetries concerns how they transform the spatial origin: rotations, inversions and reflections preserve the origin, while screw rotations and glide reflections unavoidably translate the origin by a fraction of the lattice period.\cite{Lax} If no origin exists that is simultaneously preserved, modulo lattice translations, by all symmetries in a space group, this space group is called nonsymmorphic. Despite there being more nonsymmorphic than symmorphic space groups, a nonsymmorphic insulator with nontrivial topology has yet to be found.

\begin{figure}[t]
\centering
\includegraphics[width=0.94\columnwidth]{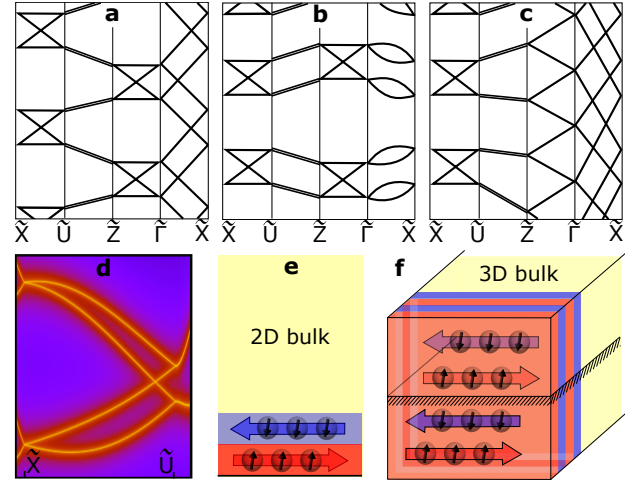}
    \caption{ Hourglass fermions and 3D quantum spin Hall effect. (a-c) Examples of the possible topologies of surface bands in a nonsymmorphic crystal. All crossings along $\tilde{Z}\tilde{\Gamma}\tilde{X}\tilde{U}$  and degeneracies along $\tilde{U}\tilde{Z}$ arise from symmetry. (a) corresponds to our material class, (b) to the trivial topology, and (c) to a nontrivial topology which may be found in other materials. (d) Hourglass fermion in KHgSb. (e) The well-known 2D quantum spin Hall effect, with a pair of spin-split modes counter-propagating on the edge. (f) 3D, doubled quantum spin Hall effect in a nutshell: two right-going surface modes with spin up, and two left-going modes with spin down. These surface modes are protected by a reflection symmetry, whose mirror plane is indicated by parallel diagonal lines.} \label{fig:cshe}
\end{figure}

In this work, we describe a new topology in band insulators that arises from fractional translations of the origin, and propose KHg$X$ ($X{=}$As,Sb,Bi) as the first material realization of its kind. The topology of KHg$X$ manifests differently on its various surfaces, depending on the spatial symmetries that are preserved on that surface. On the 010 surface, we find that the glide-mirror symmetry protects a novel surface fermion that disperses like an hourglass (see Fig.\ \ref{fig:cshe}(d)); doubly-degenerate surface bands connect one hourglass to the next in a zigzag pattern that robustly interpolates across the conduction gap in Fig.\ \ref{fig:cshe}(a). This hourglass fermion sharply contrasts with the Dirac fermions found on the surface of symmorphic topological insulators.\cite{hsieh2008} The 100 surface of KHg$X$ uniquely realizes a 3D doubled quantum spin hall effect (QSHE) with \emph{four} counter-propagating \emph{surface} modes distinguished by spin, as illustrated in Fig.\ \ref{fig:cshe}(f). Unlike the well-known 2D QSHE\cite{kane2005B,bernevig2006c,koenig2007,fu2007b,moore2007,roy2009,hsieh2008} (Fig.\ \ref{fig:cshe}(e)), the surface states of KHg$X$ are not protected by time-reversal symmetry alone, but are further stabilized by spatial symmetries. To describe the bulk topology of KHg$X$, we introduce a non-Abelian generalization of polarization that naturally describes glide-symmetric crystals, and is moreover quantized due to space-time inversion symmetry. Our work extends the well-known Abelian theory of polarization,\cite{kingsmith1993,vanderbilt1993,resta1994} which exhibits quantization due to spatial-inversion symmetry.\cite{zak1989}

\begin{figure}[H]
\centering
\includegraphics[width=0.96\columnwidth]{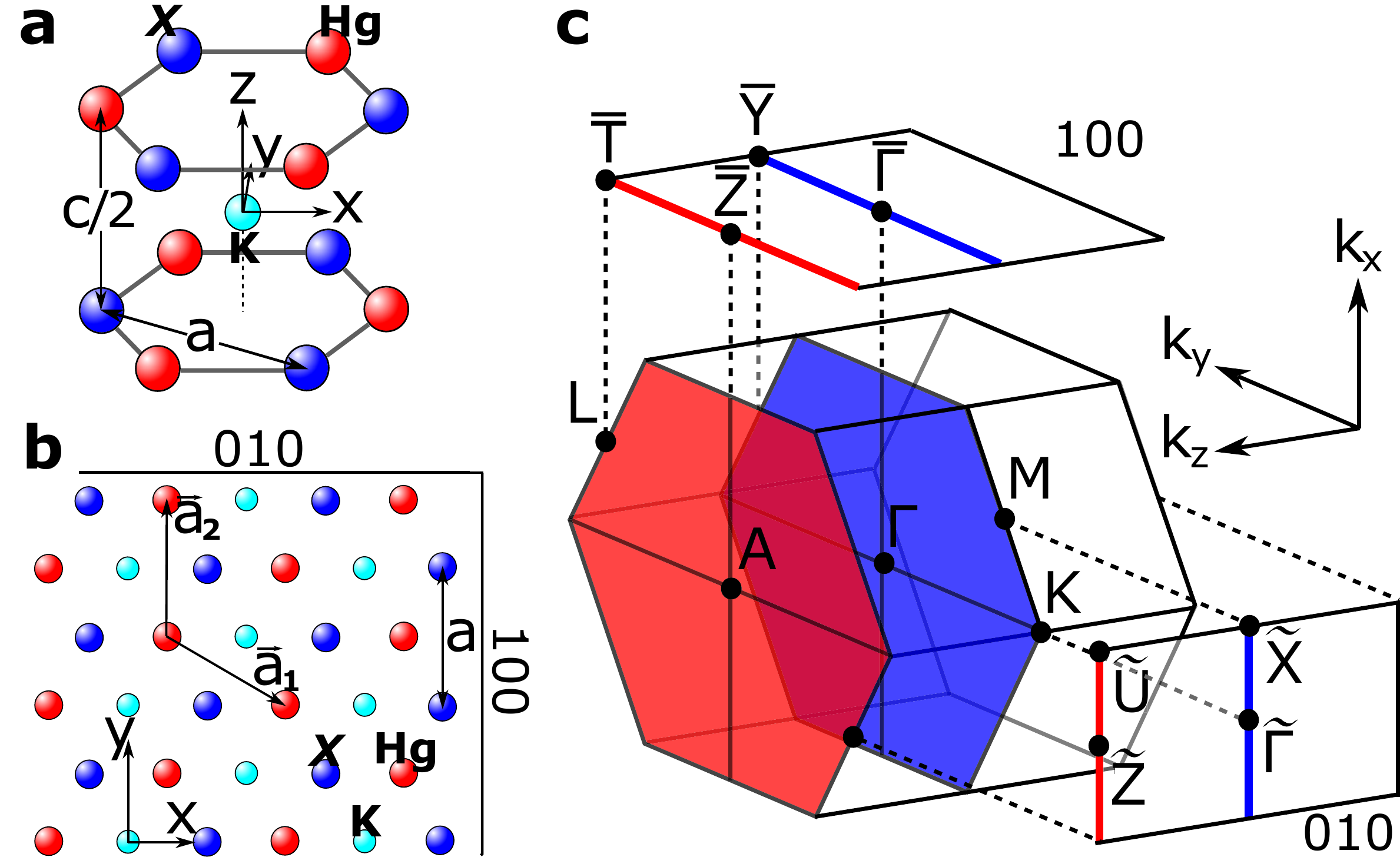}
    \caption{ Crystal structure and Brillouin zone of KHg$X$. (a) 3D view of atomic structure. The Hg (red) and $X$ (blue) ions form a honeycomb layers with AB stacking. The K ion (cyan) is located at an inversion center, which we also choose to be our spatial origin. (b) Top-down view of a truncated lattice with two surfaces labelled 010 and 100, also known respectively as ($1\bar{2}10$) and ($10\bar{1}0$) in the Miller notation. (c) Center: bulk Brillouin zone (BZ) of KHg$X$, with two mirror planes of $\bmz$ colored red and blue. Top: 100-surface BZ. Right: 010-surface BZ.  } \label{fig:structure}
\end{figure}

\noindent \textbf{{Crystal structure}} The crystal structure of KHg$X$ is illustrated in Fig.\ \ref{fig:structure}: Hg and $X$ ions form honeycomb layers with AB stacking along $\vec{z}$; between each AB bilayer sits a triangular lattice of K ions. The spatial symmetries include: (i)  an inversion ($\cali$) centered around a K ion, which we choose as our spatial origin, (ii) the screw rotation $\bar{C}_{6z}$ is a six-fold rotation about $\vec{z}$ followed by a fractional lattice translation ($t(c\vec{z}/2)$). Here and henceforth, for any transformation $g$, we denote $\bar{g} {=}  t(c\vec{z}/2) \, g$ as a product of $g$ with this fractional translation. (iii) Finally, we have the reflections $M_y: (x,y,z) {\rightarrow} (x,{-}y,z)$, $\bar{M}_z{=}t(c\vec{z}/2)M_z$ and $\bar{M}_x {=}t(c\vec{z}/2)M_x$. Among these only $\bar{M}_x$ is a glide reflection, for which the fractional translation is unremovable by a different choice of origin. Altogether, these symmetries generate the nonsymmorphic space group $D_{6h}^4(P6_3/mmc)$.\cite{exp} 


Each topological feature of KHg$X$ may be attributed to a smaller subset of the group -- on surfaces where certain bulk symmetries are lost, their associated topology is not manifest, e.g., the 100-surface symmetry is a symmorphic subgroup of $D_{6h}^4$, leading to a strikingly different bandstructure than that of the nonsymmorphic 010 surface. Our strategy is to deduce the possible topologies of the surface bands purely from representations of the surface symmetry. We then more carefully account for the bulk symmetries and their representations, as well as introduce a non-Abelian polarization to diagnose nontrivial topology in the bulk wavefunctions.

\noindent \textbf{{Surface analysis}}  Let us first discuss the 010 surface, whose group ($Pma2$) is generated by glideless $\bmz$ and glide $\bmx$. To explain the robust surface bands in Fig.\ \ref{fig:cshe}, we consider each high-symmetry line in turn: (i) At any wavevector ($\boldsymbol{k}'$) along $\tilde{Z}\tilde{U}$ ($k_z{=}\pi/c$), all bands are doubly-degenerate. Indeed, the group\cite{tinkhambook} of $\boldsymbol{k}'$ includes the antiunitary element $T\bmx$ (time reversal with a glide) which results in a Kramers-like degeneracy at each $\boldsymbol{k}'$. This follows from $(T\bmx)^2{=}T^2\bmx^2{=}t(c\vec{z})$, where the lattice translation is represented by Bloch waves as $t(c\vec{z}){=}$exp$({-}ik_z/c){=}{-}I$ along $\tilde{Z}\tilde{U}$.
 
\begin{figure}[t]
\centering
\includegraphics[width=0.96\columnwidth]{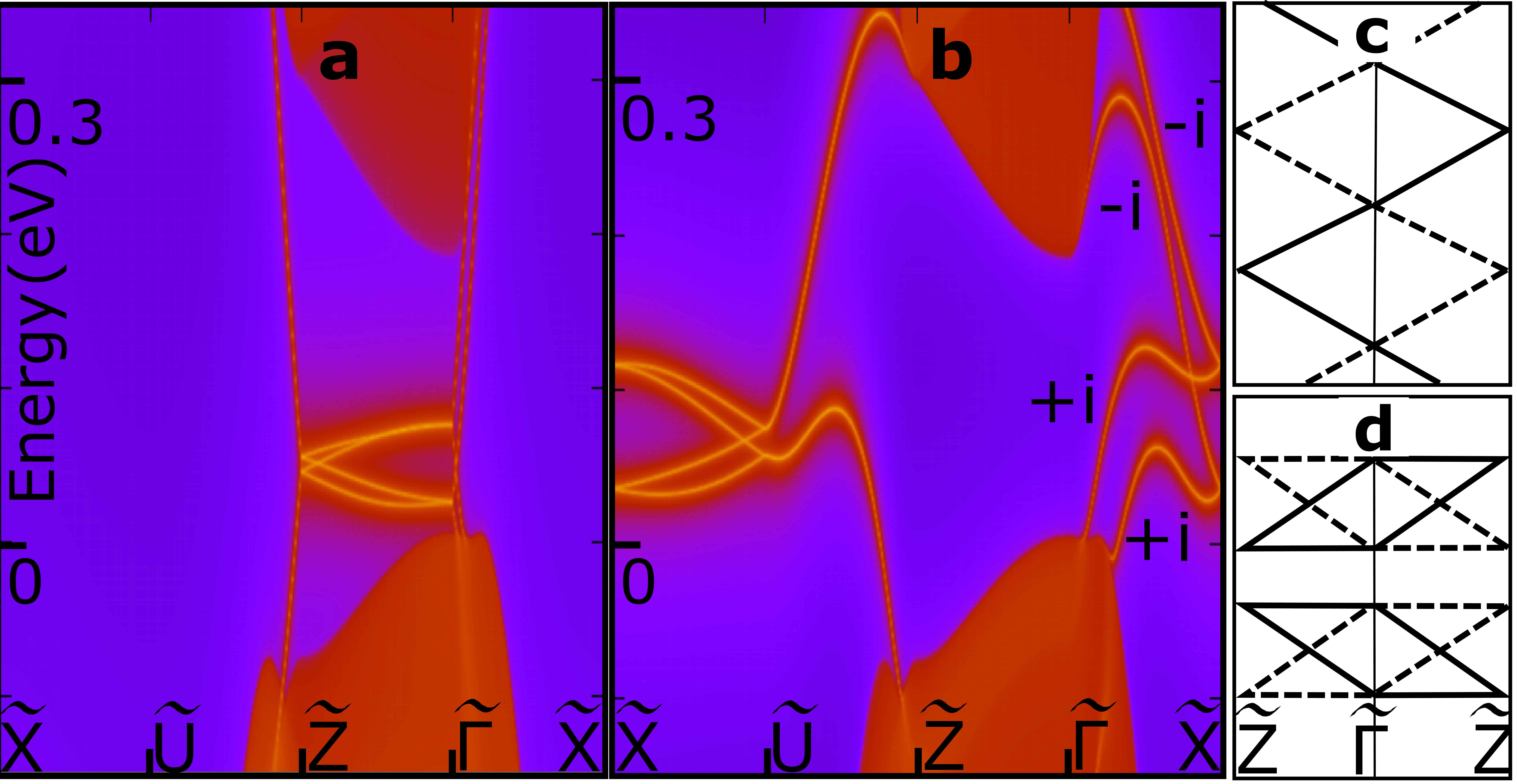}
    \caption{ The 010-surface bandstructure. The 010-surface bands of KHgSb for an ideal surface termination in (a), and with a modified surface potential in (b).\ (c-d) Possible surface topologies along $\tilde{Z}\tilde{\Gamma}\tilde{Z}$. Solid and dashed distinguish between two eigenvalue branches of $\bmx$: ${\pm}i\text{exp}({{-}ik_zc/2})$. 
} \label{fig:ysurfaceband}
\end{figure}


\noindent (ii) Along both glide-invariant lines ($\tilde{\Gamma}\tilde{Z}$  $(k_x{=}0$) and $\tilde{X}\tilde{U}$ $(k_x{=}\,\pi/\sqrt{3}a$)), bands split into quadruplets which each exhibits an internal partner-switching in the interval $k_z{\in}[0,\pi/c]$. To explain, $\bmx^2 {=}t(c\vec{z})\,\bar{E}$, with $\bar{E}$ a $2\pi$-spin rotation, implies two branches for the mirror eigenvalues: ${\pm} i$exp$({-}ik_zc/2)$. The role of time-reversal symmetry is to enforce degeneracies between complex-conjugate representations at both Kramers points, i.e., the $\bmx$ eigenvalues are paired as $\{{+}i,{-}i\}$ at $k_z{=}0$,  and either $\{{+}1,{+}1\}$ or $\{{-}1,{-}1\}$ at $k_z{=}\pi/c$. These constraints imply two topologically distinct connectivities for the surface bands. In the first (Fig.\ \ref{fig:ysurfaceband}(c)), surface bands zigzag across the conduction gap and each cusp is a Kramers doublet -- this will be elaborated as a glide-symmetric analog of the 2D QSHE\cite{kane2005A}. The second connectivity in Fig.\ \ref{fig:ysurfaceband}(d) applies to our material class: an internal partner-switching occurs within each quadruplet, resulting in an hourglass-shaped dispersion. The center of each hourglass is a robust crossing between orthogonal mirror branches, i.e., a movable but unremovable Dirac fermion in the interval $k_z{\in}[0,\pi/c]$, as exemplified by KHgSb in Fig.\ \ref{fig:cshe}(d). 



Piecing together (i) and (ii) along the {bent} line $\tilde{X}\tilde{U}\tilde{Z}\tilde{\Gamma}$, we show how a robust interpolation across the energy gap may arise. At $\tilde{Z}$ and $\tilde{U}$, there are two ways to connect hourglasses to degenerate doublets: an `hourglass flow' describes the spectral connection of all hourglasses by zigzag-connecting doublets, as drawn in the $\tilde{X}\tilde{U}\tilde{Z}\tilde{\Gamma}$ section of Fig.\ \ref{fig:cshe}(a), and further exemplified by KHgSb (in Fig.\ \ref{fig:ysurfaceband}(a)) with an ideal surface termination. To demonstrate that the surface-localized bands of KHgSb also connect with the surface-resonant bulk bands in this hourglass-flow topology, we modified the surface potential of KHgSb to push the hourglass (along $\tilde{\Gamma}\tilde{Z}$) down into the valence band; due to the proposed hourglass flow, a different hourglass is pulled down from the conduction band along $\tilde{U}\tilde{X}$ (see Fig.\ \ref{fig:ysurfaceband}(b)). In contrast, the second possible connectivity has no robust surface states (see $\tilde{X}\tilde{Z}\tilde{U}\tilde{\Gamma}$ section of Fig.\ \ref{fig:cshe}(b)). 


\noindent(iii) Along $\tilde{\Gamma}\tilde{X}$ ($k_z{=}0$), bands divide into two subspaces having either $\bmz$-eigenvalue ${+}i$ or ${-}i$, as follows from $\bmz^2{=}\bar{E}$.  As illustrated in Fig.\ \ref{fig:ysurfaceband}(b), the two chiral (anti-chiral) surface modes in the ${+}i$ (resp.\ ${-}i$) subspace may be summarized by a mirror Chern number\cite{teo2008} (MCN): $\calc_e{=}{+}2$.

\begin{figure}[H]
\centering
\includegraphics[width=0.96\columnwidth]{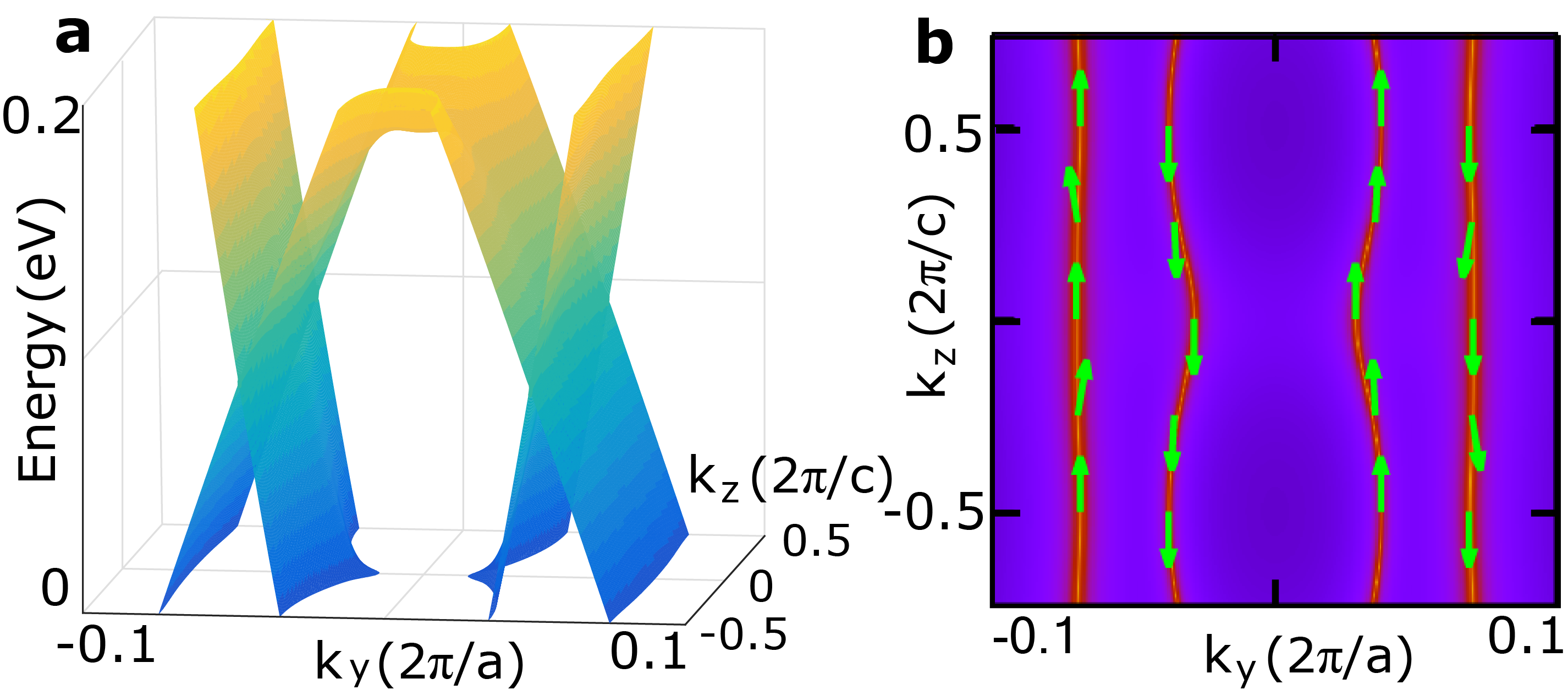}
  \caption{ The 100-surface bandstructure. (a) The 100-surface bandstructure over a momentum rectangle. While a small hybridization gap (${\sim} 1$ meV) opens for $k_z {\neq} 0$, along $k_z{=}0$ is a robust intersection  between $\bmz{=}{\pm} i$ subspaces. (b) Expectation value of spin at the Fermi level.
 } \label{fig:xsurfaceband}
\end{figure}

Since the 100 surface of KHg$X$ also preserves the glideless $\bmz$, the 100 dispersion along $\bar{\Gamma}\bar{Y}$ (see Fig.\ \ref{fig:structure}(c)) is topologically equivalent to that of the 010 along $\tilg \tilx$ -- this reflects two distinct surface projections (illustrated by blue lines in Fig.\ \ref{fig:structure}(c)) of the nontrivial MCN in the $k_z{=}0$ plane (blue plane in Fig.\ \ref{fig:structure}(c)). However, the 100 surface does not respect the glide symmetry ($\bmx$) that protects the hourglass fermions in the 010. Instead, the 100 surface modes barely disperse with $k_z$, forming the anisotropic band structure in Fig.\ \ref{fig:xsurfaceband}(a); the spin expectation value at the Fermi level is revealed in Fig.\ \ref{fig:xsurfaceband}(b). The low-energy transport properties are then described by a 3D, doubled QSHE, where at each $k_z$ we have two right-moving, spin-down, and surface-extended carriers, in combination with their time-reversed partners at ${-}k_z$. 

\begin{figure}[t]
\centering
\includegraphics[width=0.96\columnwidth]{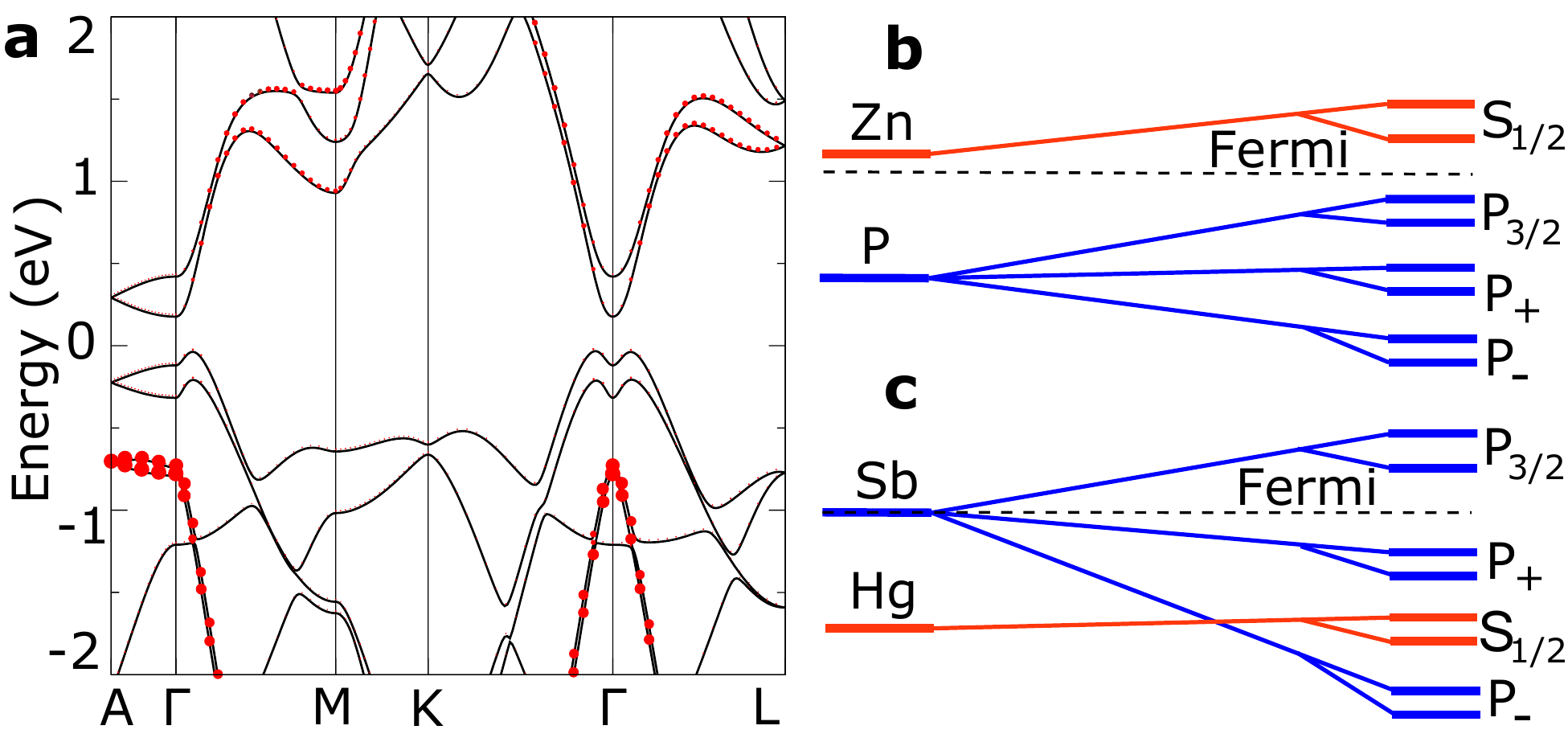}
    \caption{ Bulk bandstructure and orbital analysis. (a) The bulk bandstructure of KHgSb. The size of each red dot quantifies the weight of Hg-$s$ orbitals. (b-c) Orbital character of KZnP (top) and KHgSb (bottom) at any point along $\Gamma A$, as we vary the crystal field (CF) and spin-orbit coupling (SOC) from zero (leftmost) to their natural strengths (rightmost). The precise wavefunctions of $S_{1/2}, P_{3/2}, P_+$ and $P_-$ are clarified in the Supplemental Material. Unlike KZnP, the groundstate of KHgSb has an inverted $S_{{1}/{2}}$ quadruplet at all values of the CF and SOC, and is consequently metallic where both CF and SOC vanish.} \label{fig:bandinversion}
\end{figure}

\noindent \textbf{{Bulk analysis}} Having enumerated the possible topologies purely from an analysis of the surface symmetries, we proceed to identify which of these topologies are consistent with the bulk symmetries. In a low-energy description of KHgSb, the bulk symmetries are represented by one $s$-type quadruplet\cite{inprep} (derived from Hg) and three $p$-type quadruplets (from Sb). A $0.2$-eV bulk gap is induced by spin-orbit splitting of the $p$-type bands. Supposing electrons fill 12 of these 16 bands, two scenarios emerge: (i) If only the $p$-type bands are occupied, as exemplified by KZnP in Fig.\ \ref{fig:bandinversion}(b), then their corresponding Wannier functions will center on the P atoms from which the $p$-orbitals derive. (ii) With KHgSb, the occupied bands along $\Gamma A$ have mixed $s$- and $p$-characters (Fig.\ \ref{fig:bandinversion}(c)), which suggests that its Wannier functions center on the bond between Hg and Sb atoms. Since the 010 surface terminates to produce dangling Hb-Sb bonds (Fig.\ \ref{fig:structure}(b)), the mid-bond Wannier functions of KHgSb mutually hybridize to form surface states. Contrastingly, the on-atom Wannier functions of KZnP are unexpected to form surface states. 

\begin{figure}[t]
\centering
\includegraphics[width=0.99\columnwidth]{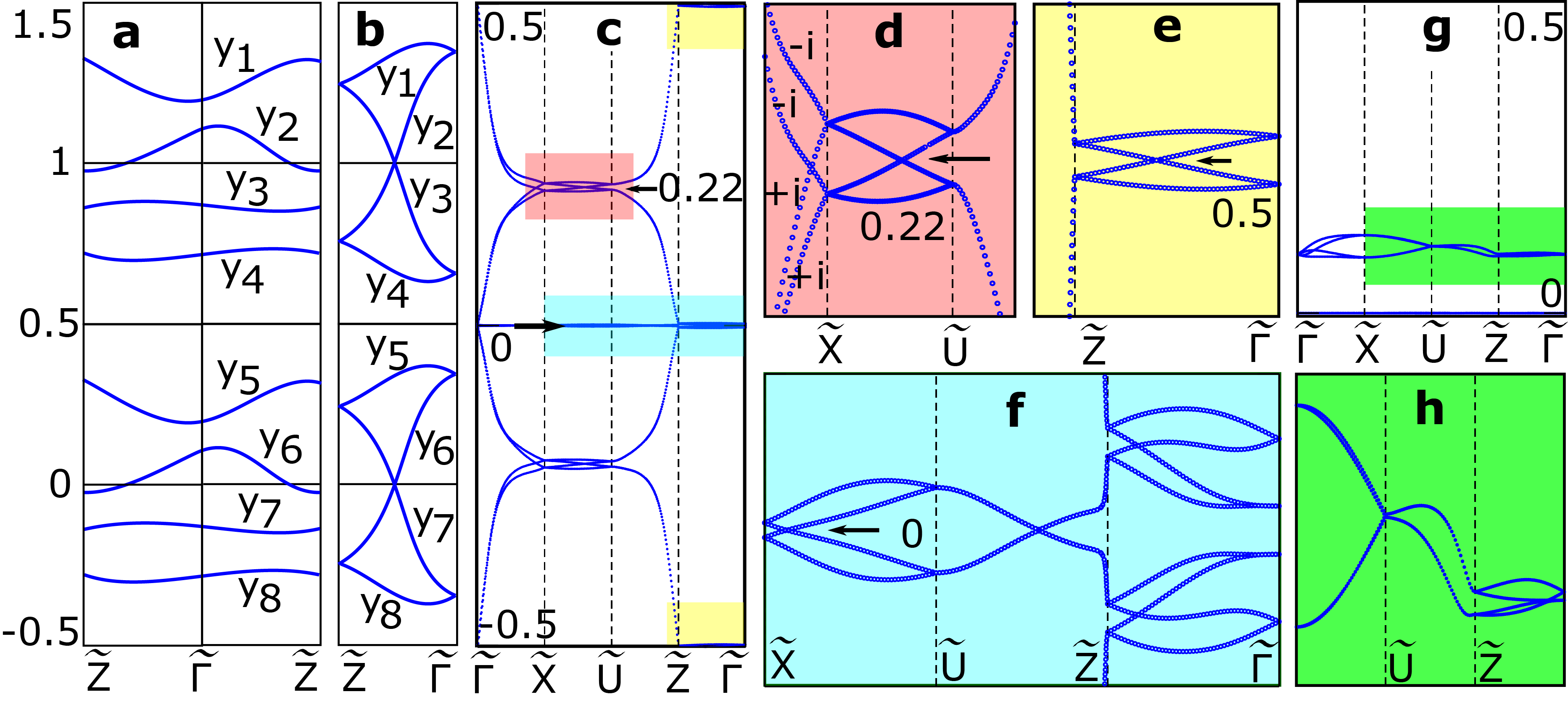}
  \caption{ Spectra of the projected-position operator $\pper \hat{y}\pper$. Comparison of the spectrum of $\pper \hat{y}\pper$ for (a) a system without any symmetry, and (b) one with time-reversal, spatial-inversion and glide symmetries. (c) Spectrum of $\pper \hat{y}\pper$ for KHgSb, with corresponding close-ups (d-f). (g) Spectrum of $\pper \hat{y}\pper$ for KZnP in half of the unit cell, with close-up (h). } \label{fig:wilson}
\end{figure}

Our intuition is justified formally by a Bloch-Wannier (BW) representation\cite{Maryam2014} of the groundstate: the $\noc$ occupied bands are represented as hybrid functions $\{|\kpar,n \rangle\ |\, n \in \{1,2,\ldots, \noc\} \}$ which maximally localize in $\vec{y}$ (as a Wannier function) but extend in $\vec{x}$ and $\vec{z}$ (as a Bloch wave with momentum $\kpar{=}(k_x,k_z)$). Each BW function is an eigenfunction of the projected-position operator $\pper\hat{y}\pper$, where $\pper$ projects to the occupied bands; the eigenvalue ($y_{\sma{n,\kpar}}$) of $\pper\hat{y}\pper$ is the center-of-mass coordinate of the BW function ($\ket{n,\kpar}$).\cite{AA2014} Due to the discrete translational symmetry of $\pper$, each $y_{\sma{n,\kpar}}$ is a mod-integer quantity representing a family of eigenfunctions related by integer translations; here, $1{\equiv}a/2$ is the translational period in $\vec{y}$ (see Fig.\ \ref{fig:structure}(a-b)). Since the spectrum of $\pper\hat{y}\pper$ can be interpolated\cite{fidk2011,ZhoushenHofstadter} to the 010-surface bandstructure (Fig.\ \ref{fig:ysurfaceband}(a)) while preserving the 010 symmetries, we expect\cite{Maryam2014} that both spectra share similar features (Fig.\ \ref{fig:wilson}): (i) degenerate doublets along $\tilz \tilu$, (ii) partner-switching quadruplets along $\tilg \tilz$ and $\tilx \tilu$, and (iii) robust crossings between orthogonal $\bmz$ subspaces (labelled by ${\pm}i$ in Fig.\ \ref{fig:wilson}(d)) along $\tilg \tilx$. 


Differences arise because the spectrum of $\pper\hat{y}\pper$ additionally encodes bulk symmetries which are spoiled by the 010 surface, e.g., while our naive surface argument allows for a glide-symmetric QSHE (i.e., zigzag connectivity in Fig.\ \ref{fig:ysurfaceband}(c)) along both $\tilg \tilz$ and $\tilx \tilu$, the out-of-surface translational symmetry rules out this scenario along $\tilx \tilu$, as shown in the Supplemental Material. A second difference originates from the bulk inversion ($\cali$) symmetry, which quantizes two invariants that have no surface analog; these invariants describe the polarization of quadruplets along the glide lines $\tilg \tilz$ and $\tilx \tilu$. Illustratively, consider in Fig.\ \ref{fig:wilson}(b) the top quadruplet, whose center-of-mass position may tentatively be defined by averaging four BW positions: $\caly_{\sma{1}}({\kpar}) {=} (1/4)\sum_{n=1}^4y_{\sma{n,\kpar}}$, with $\kpar {\in}\tilg \tilz$. Any polarization quantity should be well-defined modulo $1$, which reflects the discrete translational symmetry of the crystal. However, $\caly$ is only well-defined mod $1/4$ for quadruplet bands without symmetry, due to the integer ambiguity of each of $\{y_n |n {\in} \Z \}$. This ambiguity is illustrated in Fig.\ \ref{fig:wilson}(a) for an asymmetric insulator with four occupied bands. Only the spectrum for two spatial unit cells (with unit period) is shown, and the discrete translational symmetry ensures $y_{\sma{j,\kpar}}{=}y_{\sma{j+4l,\kpar}}{-}l$ for $j,l {\in} \Z$. Clearly the centers of mass of $\{y_1,y_2,y_3,y_4\}$ and $\{y_2,y_3,y_4,y_5\}$ differ by $1/4$ at each $\kpar$, but both choices are equally natural given level repulsion across $\tilz \tilg \tilz$. However, a unique choice for the center of mass exists if the BW bands divide into sets of four, such that within each set there are enough contact points along $\tilg \tilz$ to continuously travel between the four bands. Such a property, which we call four-fold connectivity, is illustrated in Fig.\ \ref{fig:wilson}(b) for a glide-symmetric insulator with four occupied bands ($\noc{=}4$). Here, both quadruplets $\{y_1,y_2,y_3,y_4\}$ and $\{y_5,y_6,y_7,y_8\}$ are connected, and their centers of mass differ by unity. Our definition of a mod-one center-of-mass coordinate then hinges on this four-fold connectivity which characterizes insulators with glide and time-reversal symmetries. To extend this definition to multiple quadruplets per unit cell (where integral $\noc/4{\geq}1$), let us define the net displacement of all $\noc/4$ number of connected-quadruplet centers: $\calq(\kpar)/e {=} \sum_{\sma{j{=}1}}^{\sma{\noc/4}} \caly_j(\kpar)$ mod $1$; this quantity is quantized to either $0$ or $1/2$ due to a combination of time-reversal ($T$) and spatial-inversion ($\cali$) symmetry. Indeed, $T\cali$ inverts the spatial coordinate but leaves momentum untouched: $T\cali |\kpar,n \rangle {=}|\kpar,m\rangle$ with $m {\neq} n$ and $y_{\sma{n,\kpar}} {=} {-}y_{\sma{m,\kpar}}$ mod $1$. Consequently, $T\cali : \caly_j(\kpar) {\rightarrow} \caly_{j'}(\kpar){=}{-}\caly_j(\kpar)$ mod $1$, and the only non-integer contribution to $\calq/e$ (${=}1/2$) arises if there exists a $T\cali$-invariant quadruplet ($\bar{j}$) centered at $\caly_{\bar{j}}{=}1/2{=}{-}\caly_{\bar{j}}$ mod $1$. Since each $y_{\sma{n,\kpar}}$ is a continuous function of $\kpar$, $\calq_{\sma{\kpar}}$ is constant (${\equiv}\calq_{\sma{\tilg \tilz}}$) over $\tilg \tilz$. Alternatively stated, $\calq_{\sma{\tilg \tilz}}$ is a quantized polarization invariant that characterizes the entire glide plane that projects to $\tilg \tilz$. Similarly reasoning with $\tilx \tilu$, we obtain two $\mathbb Z_2$ invariants: $\calq_{\sma{\tilg \tilz}}$ and $\calq_{\sma{\tilx \tilu}}$. 


For KHgSb, Fig.\ \ref{fig:wilson}(c) illustrates the absence (presence) of the $\caly{=}1/2$ quadruplet along $\tilx \tilu$ (resp.\ $\tilg \tilz$), leading to $\calq_{\sma{\tilx \tilu}}{=}0$ and $\calq_{\sma{\tilg \tilz}}{=}e/2$ -- this difference originates from the band inversion along $\Gamma A$ (cf.\ Fig.\ \ref{fig:bandinversion}).  Wherever $\calq_{\sma{\tilg \tilz}}{\neq}\calq_{\sma{\tilx \tilu}}$, we obtain the hourglass-flow topology exemplified in Fig.\ \ref{fig:wilson}(c). Contrastingly, Fig.\ \ref{fig:wilson}(g-h) depicts the trivial spectrum for KZnP. As initially motivated, $\calq_{\sma{\tilg \tilz}}{=}e/2$ in KHgSb indicates the mid-bond BW functions, which further hybridize to form the hourglass of Fig.\ \ref{fig:ysurfaceband}(a) when the 010 surface is terminated.

The topological distinction between KHgSb and KZnP may further be deduced by their differing quantum numbers under spatial transformations. Thus far, the most successful strategy\cite{Fu1} in finding topological materials lies in identifying centrosymmetric systems with inverted parity quantum numbers.\cite{teo2008,Hsieh2012,Xu2012,tanaka2012} For KHgSb, the parity eigenvalues of the $s$-quadruplet (recall Fig.\ \ref{fig:bandinversion}) are identical with those of any $p$-quadruplet, and therefore there is no parity inversion at any inversion-invariant momentum.\cite{zhang2011} Instead, KHgSb manifests an inversion of its eigenvalues (exp$[{-}i\pi J_z/3]$) under the screw $\bar{C}_{6z}$. $[\bar{C}_{6z},\bmz]{=}0$ implies states at $\Gamma$ can simultaneously be labelled by both operators. The $\bmz{=}{+}i$ states in the $s$-quadruplet ($p$-quadruplet) transform as $J_z{=}{-}1/2$ and $5/2$ (resp.\ $J_z{=}3/2$ and ${-}3/2$), and their inversion at $\Gamma$ results in a net angular momentum gain ($\Delta J_z{=}2$), which accompanies a quantized redistribution of Berry curvature,\cite{berry1984} i.e., $\Delta J_z$ equals the change in $\calc_e$ modulo six, as proven in the Supplemental Material. There, we further confirm $\calc_e{=}2$ by the Wilson-loop method, in accordance with our surface analysis.

\noindent \textbf{{Discussion}} Spatial symmetries have played a crucial role in the topological classification of band insulators\cite{fu2011,Hsieh2012,ChaoxingNonsymm,AAchen,chen2013,Shiozaki2014,Hoi,Varjas}; nonsymmorphic spatial symmetries are particularly useful in the classification of band semimetals\cite{connectivityMichelZak,elementaryenergybands,Young} and their Fermi-liquid analogs\cite{toplutt}, as well as in identifying topologically-ordered insulators with fractionalized excitations.\cite{nonsymmsid,nonsymmroy,watanabe} To date, all experimentally-tested topological insulators have relied on symmorphic space groups.\cite{hsieh2008,Hsieh2012,Xu2012,tanaka2012} In KHg$X$, we propose the first family of insulators with nonsymmorphic topology, in the hope of stimulating interest in an experimentally barren field. Our time-reversal-invariant theory of KHg$X$ complements previous theoretical proposals with magnetic, nonsymmorphic space groups.\cite{moore2010,ChaoxingNonsymm,chen2013,unpinned,Shiozaki2015}  We propose to characterize glide-symmetric crystals, such as KHg$X$, by a quantized polarization which depends on the non-Abelian Berry connection.\cite{wilczek1984,AA2014,berryphaseTCI} In constrast, the standard polarization relates to the Abelian Berry connection.\cite{berry1984,zak1989,kingsmith1993} Additionally, KHg$X$ uniquely exemplifies a `rotationally-inverted' insulator; a general strategy to search for such materials in all space groups is elaborated in the Supplemental Material. 

KHg$X$ represents one among many possible topologies within its space group, as illustrated in Fig.\ \ref{fig:cshe}(a-c). Which symmetry groups, other than that of KHg$X$, allow for topological surface states? We propose a criterion on the surface symmetry that applies to all known symmetry-protected surface topologies. By `symmetry-protected', we mean nonchiral surface states with vanishing Chern\cite{Haldane1988} (or mirror Chern\cite{teo2008}) numbers. Our criterion introduces the notion of connectivity within a submanifold ($\calm$) of the surface Brillouin zone, and relates to the theory of elementary energy band\cite{connectivityMichelZak,elementaryenergybands}  -- we say $\calm$ is $\cald$-fold connected if bands there divide into sets of $\cald$, such that within each set there are enough contact points in $\calm$ to continuously travel through all $\cald$ bands. If $\calm$ is a single wavevector ($\kpar$), $\cald$ coincides with the dimension of the irreducible representation at $\kpar$; $\cald$ generalizes this notion of symmetry-enforced degeneracy where $\calm$ is larger than a wavevector (e.g., a glide line). Our criterion: (a) there exist two separated submanifolds $\calm_{\sma{1}}$ and $\calm_{\sma{2}}$, with corresponding $\cald_{\sma{1}}{=}\cald_{\sma{2}}{=} fd$ ($f {\geq} 2$ and $d {\geq} 1$ are integers), and (b) a third submanifold $\calm_{\sma{3}}$ that connects $\calm_{\sma{1}}$ and $\calm_{\sma{2}}$, with corresponding $\cald_{\sma{3}}{=}d$. Almost all symmetry-protected surface topologies\cite{kane2005B,fu2011,AAchen,ChaoxingNonsymm,TCIbyrepresentation} are characterized by $\cald_{\sma{1}}{=}\cald_{\sma{2}}{=}2\cald_{\sma{3}}{=}2$, with $\calm_{\sma{1}}$ and $\calm_{\sma{2}}$ two high-symmetry wavevectors connected by a curve $\calm_{\sma{3}}$, e.g., the edge of the QSH insulator\cite{kane2005A} is characterized by two Kramers-degenerate momenta (hence $\cald_{\sma{1}}{=}\cald_{\sma{2}}{=}2$) connected by  a curve with trivial degeneracy ($\cald_{\sma{3}}{=}1$) -- these constraints allow for a Kramers-partner-switching dispersion.\cite{kane2005B} In this work, the surface symmetry $Pma2$ is characterized by two glide lines ($\calm_{\sma{1}}{=}\tilg \tilz,\calm_{\sma{2}}{=}\tilx \tilu$) with hourglass bandstructures ($\cald_{\sma{1}}{=}\cald_{\sma{2}}{=}4$), and a glideless mirror line ($\calm_{\sma{3}}{=}\tilz \tilu$) with doubly-degenerate bands ($\cald_{\sma{3}}{=}2$). Previous studies of magnetic systems\cite{unpinned,Shiozaki2015,lu2015,Varjas} have established a $\mathbb Z_2$ topology with $\cald_{\sma{1}}{=}\cald_{\sma{2}}{=}2\cald_{\sma{3}}{=}2$, where $\calm_{\sma{1}}$ and $\calm_{\sma{2}}$ are also parallel glide lines. Our surface-centric criterion for nontrivial topology is sometimes over-predictive because it neglects bulk symmetries spoilt by the surface -- a fully-predictive methodology involves the representation theory of Wilson loops and the new notion of a cohomological insulator.\cite{inprep} Finally, an exciting direction for future research lies in gapping the hourglass fermion with magnetism and superconductivity.\cite{fu2008} \\

\noindent \textbf{Acknowledgments} We thank Chen Fang, Daniel P. Arovas, Jian Li, Lukas Muechler and Xi Dai for valuable discussions. This work was supported by NSF CAREER DMR-095242, ONR-N00014\text{-}11\text{-}1-0635, ARO MURI on topological insulators, grant W911NF-12-1-0461,  NSF-MRSEC DMR-1420541, Packard Foundation, Keck grant, ``ONR Majorana Fermions'' 25812\text{-}G0001\text{-}10006242\text{-}101, and Schmidt fund 23800\text{-}E2359\text{-}FB625. During the refereeing stages of this work, AA was supported by the Yale Fellowship in Condensed Matter Physics.\\

\noindent \textbf{Author contributions} ZW and AA contributed equally to this work. AA, ZW and BAB performed theoretical analysis; ZW discovered the KHg$X$ material class and performed the first-principles calculations; RJC provided several other material suggestions.

\bibliography{nonsymm_refs}

\begin{widetext}

\ \\
\newpage

\newpage

\begin{appendix}

Organization of the Supplemental Material:\\

\noi{\ref{app:methodsdft}} We briefly describe the methods used in our first-principles calculations, as well as reveal results which are not available in the main text: (i) our calculations demonstrate a larger family of materials belong to the same topological class as KHgSb, and (ii) we also show more details of the 100- and 010-surface bands; the latter, in particular, exhibits a Lifshitz transition. \\

\noi{\ref{app:reviewtightbind}} We review the tight-binding method, emphasizing the constraints imposed by space-time symmetries. Notations are introduced which will be employed in the remaining appendices. \\

\noi{\ref{app:kp}} The general symmetry discussion of the previous Appendix is now applied to KHgSb. A detailed orbital analysis is presented, as well as effective Hamiltonians that clarify the nature of the band inversion.\\

\noi{\ref{app:bulkinvariants}} We elaborate on the diagnosis of topological invariants from bulk wavefunctions. Our diagnosis tool is the Wilson-loop operator, which we is synonymous with the projected-position operator. Through the Wilson loop, we propose an efficient method to diagnose mirror Chern numbers in time-reversal-invariant systems, and apply our method to our material class. We also prove that some topologies cannot exist on certain mirror planes.\\

\noi{\ref{rotationalinversion}} We outline a general strategy to find topological materials which are rotationally inverted, as well as provide a detailed case study for our material class.

\section{First-principles study: methods and results} \label{app:methodsdft} 

\subsection{Bulk bandstructure}

We perform electronic structure calculations within density-functional theory (DFT) as implemented in the Vienna \textit{ab initio} simulation package,\cite{kresse1993} and use the core-electron projector-augmented-wave basis in the generalized-gradient method.\cite{perdew1996} Spin-orbital coupling (SOC) is accounted for self-consistently. The cutoff energy for wave-function expansion is 500 eV, and the k-point sampling grid is 16$\times$16$\times$8. To date, KHgSb and KHgAs are only known to crystallize with $D_{6h}^4$ symmetry, and their bandstructures are calculated with the lattice constants from Ref.\ \onlinecite{exp}; however, the lack of experimental parameters for KHgBi requires that we numerically optimize its lattice constants. As shown in Fig.\ \ref{fig:AsBi}, the bandstructures of KHgAs and KHgBi show a similar band inversion as in KHgSb, indicating that they belong to the same topological class. Besides, our first-principles calculations also show RbHgX and NaHgSb belong to the same topological class if they crystallize with $D_{6h}^4$ symmetry.

\begin{figure}[H]
\centering
\includegraphics[width=8 cm]{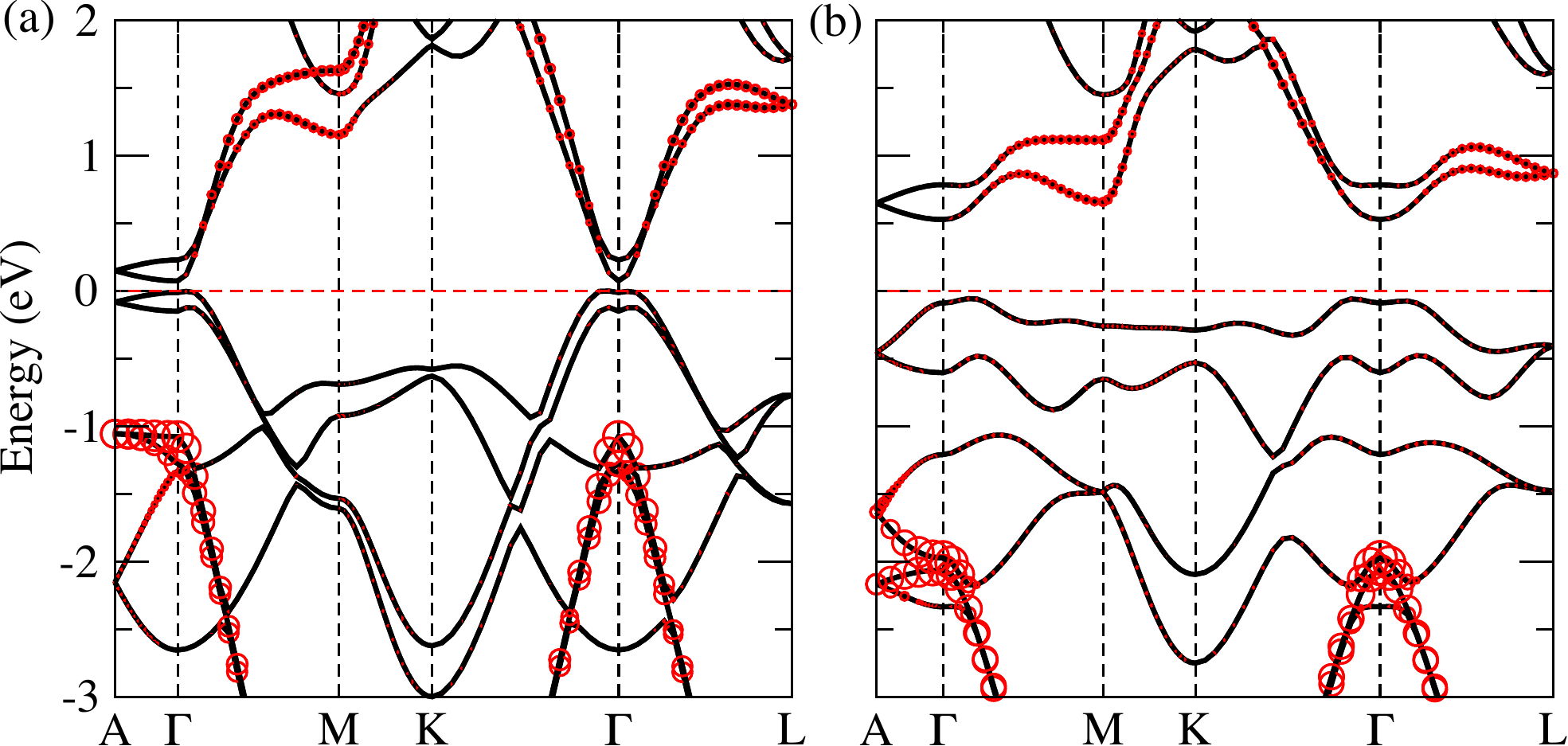}
    \caption{(a) Bulk bandstructure of KHgAs shows a gap of $0.1$ eV. (b) KHgBi with a gap of $0.5$ eV. For both (a-b), the size of each red dot quantifies the weight of Hg-$s$ orbitals.} \label{fig:AsBi}
\end{figure}

\subsection{Surface bandstructure}

To obtain the surface bandstructure, we first constructed the maximally-localized Wannier functions (WF's) from the first-principles bulk wavefunctions. These WF's were employed in a surface Green's function calculation, for a semi-infinite system.\cite{zhang2009} The result of such a calculation is shown in Fig.\ \ref{fig:Lifshitz}, for both 010 and 100 surfaces. We briefly comment on the 001 surface, whose symmetry group ($C_{3v}^{\sma{(a)}}$)\cite{AAchen} is generated by the rotation $C_{3z}$ and the symmorphic reflection $M_y$; here, the six-fold screw rotation ($\bar{C}_{6z}$) is spoilt because the fractional translation in $\bar{C}_{6z}$ is orthogonal to the surface, but the product of two six-fold screw rotations is, modulo an integral-lattice translation, a symmorphic three-fold rotation $C_{3z}$ that is a symmetry of the 001 surface. Due to the triviality of the $M_y$-Chern number and of all invariants protected by time-reversal symmetry,\cite{Fuz2,Fu1,fu2007b} the 001 is completely absent of robust surface states.\\

\begin{figure}[H]
\centering
\includegraphics[width=8.6 cm]{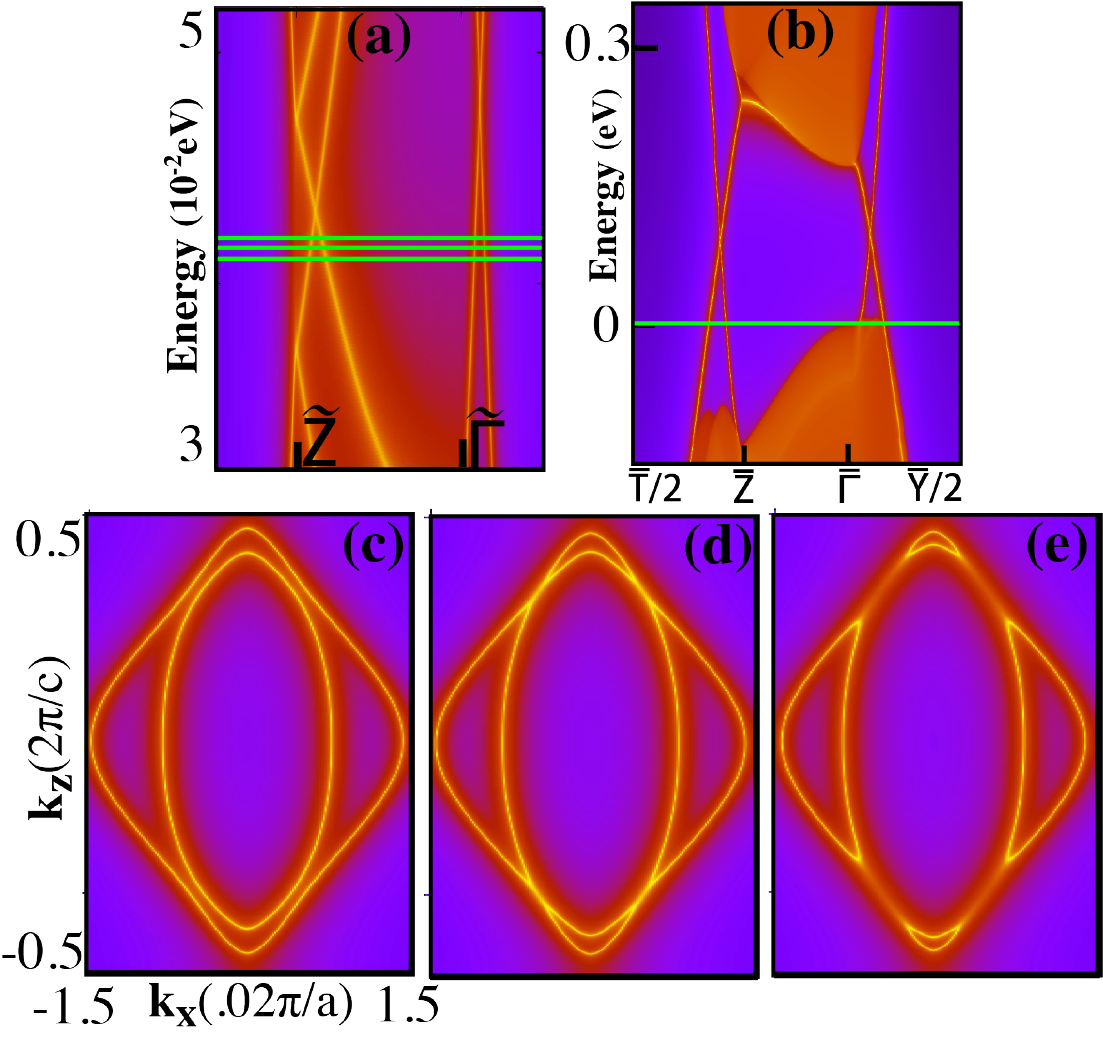}
    \caption{(a) is a close-up of the hourglass fermion on the 010 surface, and the three green lines indicate the energy range where a Lifshitz transition occurs. (b) 100-surface bandstructure along a high-symmetry line. (c-e) Constant-energy contours at the three energies indicated in (a).} \label{fig:Lifshitz}
\end{figure}

\section{Review of symmetries in the tight-binding method} \label{app:reviewtightbind}

We introduce our notation for the tight-binding method in Sec.\ \ref{sec:general}, then consider the effects of spatial symmetries and space-time symmetries in Sec.\ \ref{app:rep} and \ref{sec:spacetimetight} respectively.

\subsection{General remarks on the tight-binding method} \label{sec:general}

In the tight-binding method, the Hilbert space is reduced to a finite number ($n_{tot}$) of \low orbitals $\varphi_{\boldsymbol{R},\alpha}$, for each unit cell labeled by the Bravais lattice (BL) vector $\boldsymbol{R}$.\cite{slater1954,goringe1997,lowdin1950} In Hamiltonians with discrete translational symmetry, our basis vectors are
\begin{align} \label{basisvec}
\phi_{\boldsymbol{k}, \alpha}(\boldsymbol{r}) = \tfrac{1}{\sqrt{N}} \sum_{\boldsymbol{R}} e^{i\boldsymbol{k} \cdot (\boldsymbol{R}+\boldsymbol{r_{\alpha}})} \pdg{\varphi}_{\boldsymbol{R},\alpha}(\boldsymbol{r}-\boldsymbol{R}-\boldsymbol{r_{\alpha}}); \;\;\;\; \alpha=1,\ldots, n_{tot},
\end{align}
which are periodic in lattice translations $\boldsymbol{R}$. $\boldsymbol{k}$ is a crystal momentum, $N$ is the number of unit cells, $\alpha$ labels the \low orbital, and $\boldsymbol{r_{\alpha}}$ denotes the position of the orbital $\alpha$ as measured from the origin in each unit cell. The tight-binding, Bloch Hamiltonian is defined as 
\begin{align} \label{tightbindingbloch}
H(\boldsymbol{k})_{\alpha \beta} = \int d^dr\,\phi^*_{\boldsymbol{k},\alpha}(\boldsymbol{r}) \,\hat{H} \,\phi_{\boldsymbol{k},\beta}(\boldsymbol{r}), 
\end{align}
where $\hat{H}$ is the single-particle Hamiltonian. The energy eigenstates are labeled by a band index $j$, and defined as $\psi_{j,\boldsymbol{k}}(\boldsymbol{r}) = \sum_{\alpha=1}^{n_{tot}}  \,u_{j,\boldsymbol{k}}(\alpha)\,\phi_{\boldsymbol{k}, \alpha}(\boldsymbol{r})$, where  
\begin{align}
\sum_{\beta=1}^{n_{tot}} H(\boldsymbol{k})_{\ab} \,u_{j,\boldsymbol{k}}(\beta)  = \varepsilon_{j,\boldsymbol{k}}\,u_{j,\boldsymbol{k}}(\alpha).
\end{align}
We employ the braket notation:
\begin{align} \label{energyeigen}
H(\boldsymbol{k})\,\ket{u_{j,\boldsymbol{k}}} = \varepsilon_{j,\boldsymbol{k}}\,\ket{u_{j,\boldsymbol{k}}}.
\end{align}
Due to the spatial embedding of the orbitals, the basis vectors $\phi_{\boldsymbol{k},\alpha}$ are generally not periodic under $\boldsymbol{k} \rightarrow \boldsymbol{k}+\boldsymbol{G}$ for a reciprocal lattice (RL) vector $\boldsymbol{G}$. This implies that the tight-binding Hamiltonian satisfies:
\begin{align} \label{aperiodic}
H(\boldsymbol{k}+\boldsymbol{G}) = V(\boldsymbol{G})^{\mo}\,H(\boldsymbol{k})\,V(\boldsymbol{G}),
\end{align}
where $V(\boldsymbol{G})$ is a unitary matrix with elements: $[V(\boldsymbol{G})]_{\ab} = \delta_{\ab}\,e^{i\boldsymbol{G}\cdot \boldsymbol{r_{\alpha}}}$.\\

We are interested in Hamiltonians with a spectral gap that is finite throughout the Brillouin zone (BZ), such that we can distinguish occupied from empty bands. Let $P$ project to the occupied bands as
\begin{align} \label{periodP}
P= \sum_{\bk \in BZ} P(\bk) \ins{and} P(\bk) = \sum_{n =1}^{\noc} \ket{u_{n,\bk}}\bra{u_{n,\bk}} = V(\bG)\,P(\bk+\bG)\,V(\bG)^{\mo}.
\end{align}

\subsection{Effect of spatial symmetries on the tight-binding Hamiltonian} \label{app:rep}

Let us denote a spatial transformation by $\pdg{g}_{\bdelta}$, which transforms real-space coordinates as $\br \rightarrow D_g \br + \bdelta$, where $D_g$ is the orthogonal matrix representation of the point-group transformation $g$ in $\R^d$.  Nonsymmorphic space groups contain symmetry elements where $\bdelta$ is a rational fraction\cite{Lax} of the lattice period; in a symmorphic space group, an origin can be found where $\bdelta=0$ for all symmetry elements. The purpose of this Section is to derive the constraints of $\pdg{g}_{\bdelta}$ on the tight-binding Hamiltonian. First, we clarify how $\pdg{g}_{\bdelta}$ transforms the creation and annihilation operators. We define the creation operator for a \low function\cite{slater1954,goringe1997,lowdin1950} ($\varphi_{\alpha}$) at Bravais lattice vector $\boldsymbol{R}$ as $\dg{c}_{\alpha}(\boldsymbol{R}+\boldsymbol{r_{\alpha}})$. From (\ref{basisvec}), the creation operator for a Bloch-wave-transformed \low orbital $\phi_{\boldsymbol{k},\alpha}$ is
\begin{align} \label{blochbasis}
\dg{c}_{\boldsymbol{k}, \alpha} = \frac{1}{\sqrt{N}} \sum_{\boldsymbol{R}} \,e^{i\boldsymbol{k} \cdot (\boldsymbol{R} + \boldsymbol{r_{\alpha}})}\,\dg{c}_{\alpha}(\boldsymbol{R}+\boldsymbol{r_{\alpha}}); \myspace \alpha =1,\ldots,n_{tot}.
\end{align}
A Bravais lattice (BL) that is symmetric under $\pdg{g}_{\bdelta}$ satisfies two conditions:\\

\noi{i}  for any BL vector $\boldsymbol{R}$, $D_g\boldsymbol{R}$ is also a BL vector: 
\begin{align} \label{pgscond1}
\forall \boldsymbol{R} \in \text{BL}, \;\; D_g\boldsymbol{R} \in \text{BL}.
\end{align}
\noi{ii} If $\pdg{g}_{\bdelta}$ transforms an orbital of type $\alpha$ to another of type $\beta$, then $D_g(\boldsymbol{R}+\boldsymbol{r_{\alpha}})+\bdelta$ must be the spatial coordinate of an orbital of type $\beta$. To restate this formally, we define a matrix $U_{\sma{g\bdelta}}$ such that the creation operators transform as
\begin{align}
\pgdel\;:\;\dg{c}_{\alpha}(\boldsymbol{R}+\boldsymbol{r_{\alpha}}) \; \longrightarrow \; \dg{c}_{\beta}\big(\,D_g\boldsymbol{R}+\boldsymbol{R}^{\sma{g \bdelta}}_{\sma{\beta \alpha}}+\boldsymbol{r_{\beta}}\,\big) \,[U_{\sma{g\bdelta}}]_{\beta \alpha},
\end{align}
with $\boldsymbol{R}^{\sma{g \bdelta}}_{\sma{\beta \alpha}} \equiv D_g\boldsymbol{r_{\alpha}}+\bdelta -\boldsymbol{r_{\beta}}$. Then\begin{align} \label{pgscond2}
[U_{\sma{g\bdelta}}]_{\beta \alpha} \neq 0 \imp  \boldsymbol{R}^{\sma{g \bdelta}}_{\sma{\beta \alpha}} \in \text{BL}. 
\end{align}
Explicitly, the nonzero matrix elements are given by
\begin{align}
[U_{\sma{g\bdelta}}]_{\beta \alpha} = \sum_{s,s'} \int d^d r\; \varphi_{\beta}^*(\br,s')\; [D^{(1/2)}_g]_{s's} \;\varphi_{\alpha}(D_g^{-1}\br,s),
\end{align}
where $\varphi_{\sma{\alpha}}$ is a spinor with spin index $s$, and $D_{\sma{g}}^{\sma{(1/2)}}$ represents $\pdg{g}_{\bdelta}$ in the spinor representation. \\

For fixed $\pdg{g}_{\bdelta},\alpha$ and $\beta$, such that $[U_{\sma{g\bdelta}}]_{\beta \alpha} \neq 0$, the mapping $\calt^{\sma{g \bdelta}}_{\sma{\beta \alpha}}: \boldsymbol{R} \rightarrow \boldsymbol{R}^{\sma{g \bdelta}}_{\sma{\beta \alpha}} \in BL$ is bijective. Applying (\ref{blochbasis}), (\ref{pgscond1}), (\ref{pgscond2}), the orthogonality of $D_g$ and the bijectivity of $\calt^{\sma{g \bdelta}}_{\sma{\beta \alpha}}$, the Bloch basis vectors transform as
\begin{align} \label{blochspatial}
\pgdel\; : \; \dg{c}_{\boldsymbol{k},\alpha} \; \longrightarrow &\; \frac{1}{\sqrt{N}} \sum_{\boldsymbol{R}} \,e^{i\boldsymbol{k} \cdot (\boldsymbol{R} + \boldsymbol{r_{\alpha}})}\, \dg{c}_{\beta}\big(\,D_g\boldsymbol{R}+\boldsymbol{R}^{\sma{g \bdelta}}_{\sma{\beta \alpha}}+\boldsymbol{r_{\beta}}\,\big) \,[U_{\sma{g\bdelta}}]_{\beta \alpha}\notag \\
\eq e^{-i (D_g \boldsymbol{k}) \cdot \bdelta } \frac{1}{\sqrt{N}} \sum_{\boldsymbol{R}} \,e^{i[D_g\boldsymbol{k}] \cdot [D_g(\boldsymbol{R} + \boldsymbol{r_{\alpha}})+\bdelta]}\, \dg{c}_{\beta}\big(\,D_g\boldsymbol{R}+\boldsymbol{R}^{\sma{g \bdelta}}_{\sma{\beta \alpha}}+\boldsymbol{r_{\beta}}\,\big) \,[U_{\sma{g\bdelta}}]_{\beta \alpha}\notag \\
\eq e^{-i (D_g \boldsymbol{k}) \cdot \bdelta } \frac{1}{\sqrt{N}} \sum_{\boldsymbol{R}} \,e^{i[D_g\boldsymbol{k}] \cdot [D_g\boldsymbol{R}+\boldsymbol{R}^{\sma{g \bdelta}}_{\sma{\beta \alpha}}+\boldsymbol{r_{\beta}}]}\, \dg{c}_{\beta}\big(\,D_g\boldsymbol{R}+\boldsymbol{R}^{\sma{g \bdelta}}_{\sma{\beta \alpha}}+\boldsymbol{r_{\beta}}\,\big) \,[U_{\sma{g\bdelta}}]_{\beta \alpha}\notag \\
\eq e^{-i (D_g \boldsymbol{k}) \cdot \bdelta } \frac{1}{\sqrt{N}} \sum_{\boldsymbol{R}'} \,e^{i[D_g\boldsymbol{k}] \cdot [\boldsymbol{R}'+\boldsymbol{r_{\beta}}]}\, \dg{c}_{\beta}\big(\,\boldsymbol{R}'+\boldsymbol{r_{\beta}}\,\big) \,[U_{\sma{g\bdelta}}]_{\beta \alpha}\notag \\
\eq  e^{-i (D_g \boldsymbol{k}) \cdot \bdelta } \dg{c}_{D_g\boldsymbol{k},\beta}\,[U_{\sma{g\bdelta}}]_{\beta \alpha}.
\end{align}
This motivates a definition of the operator
\begin{align} \label{actbloch}
\pdg{\hat{g}}_{\bdelta}(\bk) \equiv e^{-i (D_g \boldsymbol{k}) \cdot \bdelta }\, U_{\sma{g\bdelta}},
\end{align}
which acts on Bloch wavefunctions ($|u_{n,\bk}\rangle$) as
\begin{align} \label{act}
\pgdel\; : \;\ket{u_{n,\bk}} \;\longrightarrow \; \pdg{\hat{g}}_{\bdelta}(\bk)\,\ket{u_{n,\bk}}.
\end{align}
The operators $\{\pdg{\hat{g}}_{\bdelta}(\bk)\}$ form a representation of the space-group algebra\cite{Lax} in a basis of Bloch-wave-transformed \low orbitals; we call this the \low representation. If the space group is nonsymmorphic, the nontrivial phase factor exp$(-i D_g \bk \cdot \bdelta)$ in $\hatgdel(\bk)$ encodes the effect of the fractional translation, i.e., the momentum-independent matrices $\{U_{\sma{g\bdelta}}\}$ by themselves form a representation of a point group.

\begin{figure}[H]
\centering
\includegraphics[width=8 cm]{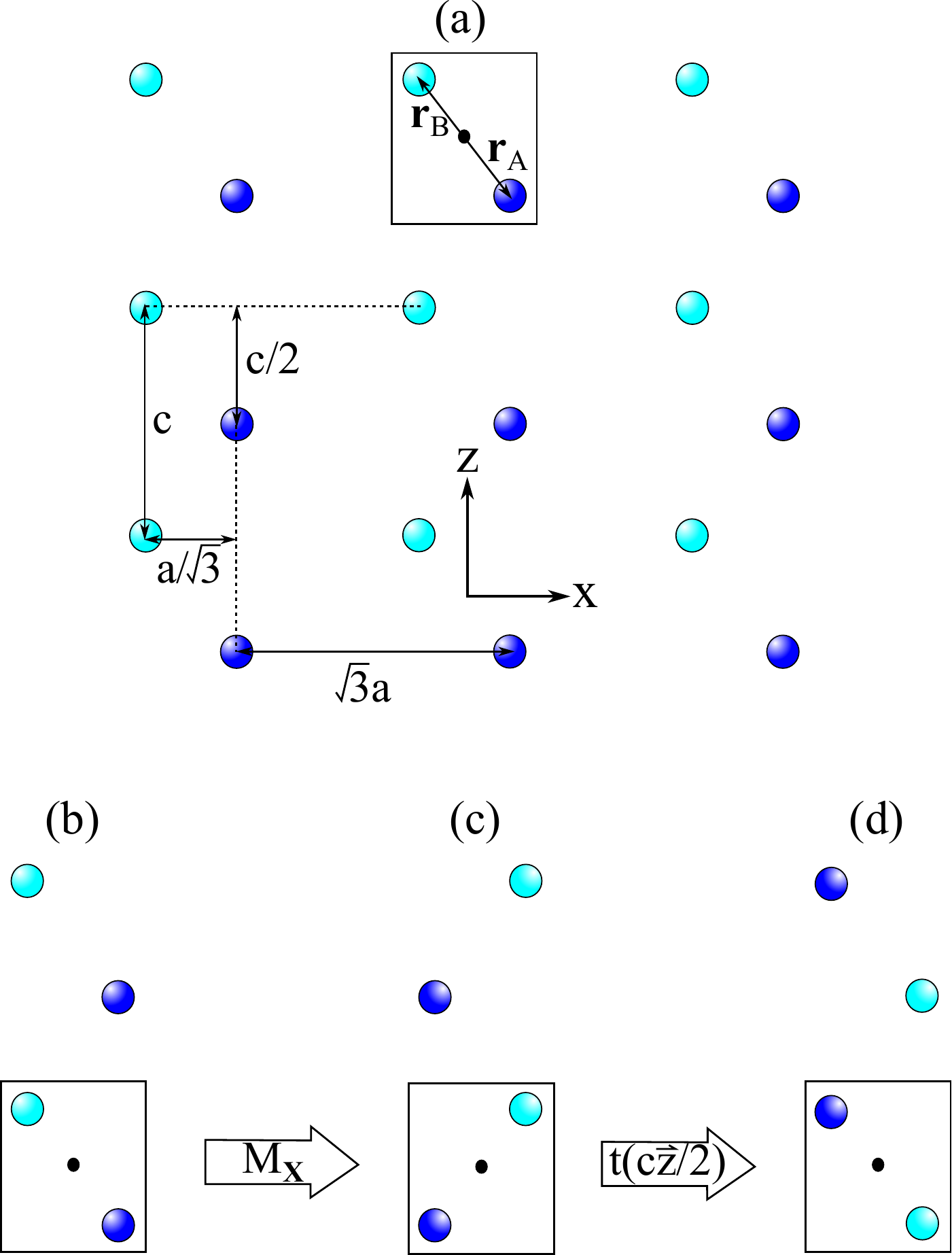}
    \caption{(a) Simple example of a 2D nonsymmorphic crystal. The two sublattices are colored respectively colored dark blue and cyan. (b-d) illustrate the effect of a glide reflection. } \label{fig:nonsymmorphic_ex}
\end{figure}

To exemplify this abstract discussion, we analyze a simple 2D nonsymmorphic crystal in Fig.\ \ref{fig:nonsymmorphic_ex}. As delineated by a square, the unit cell comprises two \emph{same} atoms labelled by subcell coordinates $A$ and $B$, and the spatial origin is chosen at their midpoint, such that $\br_{\sma{A}}=a\vec{x}/\sqrt{3} - c\vec{z}/2 =-\br_{\sma{B}}$, as shown in Fig.\ \ref{fig:nonsymmorphic_ex}(a). The symmetry group ($Pma2$) of this lattice is generated by the elements ${\bmx}$ and $\bmz$, where in the former we first reflect across $\vec{x}$ ($g=M_x$) and then translate by $\bdelta=c\vec{z}/2$. Similarly, $\bmz$ is shorthand for a $z{\rightarrow}{-}z$ reflection followed by a translation by $\bdelta=c\vec{z}/2$. Let us represent these symmetries with spin-doubled $s$ orbitals on each atom. Choosing our basis to diagonalize $S_z$, 
\begin{align}
&\bmx\;:\; \begin{cases} \dg{c}_{\sma{A,S_z}}(\bR+\br_{\sma{A}})  \; \longrightarrow \; -i\dg{c}_{\sma{B,-S_z}}(D_x\bR+\br_{\sma{B}}), \\
\dg{c}_{\sma{B,S_z}}(\bR+\br_{\sma{B}})\; \longrightarrow \;  -i\dg{c}_{\sma{A,-S_z}}(D_x\bR+c\vec{z}+\br_{\sma{A}}), \end{cases}
\end{align}
where $D_x(x,z)^t = (-x,z)^t$, and in the second mapping, we have applied $\bR^{\sma{x,c\vec{z}/2}}_{\sma{AB}} = D_x\br_{\sma{B}}+ c\vec{z}/2 - \br_{\sma{A}} = c\vec{z}$. It is useful to recall here that a reflection is the product of an inversion with a two-fold rotation about the reflection axis: $M_{\sma{j}}=\cali\, C_{\sma{2j}}$ for $j \in \{x,z\}$. Consequently, $\bmx \propto C_{\sma{2x}}$ flips $S_{\sma{z}} \rightarrow -S_{\sma{z}}$. In the basis of Bloch waves,
\begin{align}
&\bmx\;:\;  \dg{c}_{k,\alpha}\; \longrightarrow \;  e^{-ik_zc/2}\,\dg{c}_{D_x\bk,\beta}\,[U_{\bmx}]_{\beta \alpha} \ins{with} U_{\bmx} = -i\,\tx\, \sx. 
\end{align}
Here, we have employed $\tz=+1$ ($-1$) for subcell $A$ ($B$) and $\sz=+1$ for spin up in $\vec{z}$. A similar analysis for the other reflection ($\bmz \propto C_{\sma{2z}} \propto \text{exp}[-iS_z\pi]$) leads to
\begin{align}
&\bmz\;:\; \begin{cases} \dg{c}_{\sma{A,S_z}}(\bR+\br_{\sma{A}})  \; \longrightarrow \; -i\,\text{sign}[S_z]\,\dg{c}_{\sma{A,S_z}}(D_z\bR+c\vec{z}+\br_{\sma{A}}), \\
\dg{c}_{\sma{B,S_z}}(\bR+\br_{\sma{B}})\; \longrightarrow \;  -i\,\text{sign}[S_z]\,\dg{c}_{\sma{B,S_z}}(D_z\bR+\br_{\sma{B}}), \end{cases}
\end{align}
with $D_z(x,z)^t = (x,-z)^t$, and in the basis of Bloch-wave-transformed \low orbitals,
\begin{align}
&\bmz\;:\;  \dg{c}_{\bk,\alpha}\; \longrightarrow \;  e^{-ik_zc/2}\,\dg{c}_{D_z\bk,\beta}\,[U_{\bmz}]_{\beta \alpha} \ins{with} U_{\bmz} = -i\,\sz. 
\end{align}
To recapitulate, we have derived $\{\hatgdel\}$ as 
\begin{align}
\hat{\bar{M}}_x(\bk) =  -i\,e^{-ik_zc/2}\,\tx\,\sx \ins{and} \hat{\bar{M}}_z(\bk) =  -i\,e^{-ik_zc/2}\,\sz, 
\end{align}
which should satisfy the space-group algebra for $Pma2$, namely that
\begin{align}
\bmx^2 = \bar{E}\,t(c\vec{z}),\;\;\;\; \bmz^2 = \bar{E},\ins{and} \bmz\,\bmx = \bar{E}\,t(-c\vec{z})\,\bmx\,\bmz,
\end{align}
where $\bar{E}$ denotes a $2\pi$ rotation and $t(c\vec{z})$ a translation. Indeed, when acting on Bloch waves with momentum $\bk$,
\begin{align}
\hat{\bar{M}}_x&(D_x\bk)\;\hat{\bar{M}}_x(\bk) = -e^{-ik_zc},\;\;\;\; \hat{\bar{M}}_z(D_z\bk)\;\hat{\bar{M}}_z(\bk) = -I, \ins{and}\notag \\
&\hat{\bar{M}}_{z}(D_x\bk)\,\hat{\bar{M}}_x(\bk) = -e^{-ik_zc}\,\hat{\bar{M}}_x(D_z\bk)\,\hat{\bar{M}}_z(\bk).
\end{align}
Finally, we verify that the momentum-independent matrices $\{U_{\sma{g\bdelta}}\}$ form a representation of the double point group $C_{2v}$, whose algebra is simply
\begin{align} \label{symmorphicalgebratwo}
M_x^2=M_z^2= \bar{E} \ins{and} M_z\,M_x =\bar{E}\,M_x\,M_z.
\end{align}
A simple exercise leads to
\begin{align}
U_{\bmx}^2=U_{\bmz}^2=-I \ins{and} \{U_{\bmx},U_{\bmz}\}=0.
\end{align}
The algebras of $C_{2v}$ and $Pma2$ differ only in the additional elements $t({\pm} c \vec{z})$, which in the \low representation ($\{\hatgdel(\bk)\}$) is accounted for by the phase factors exp$(-ik_zc/2)$.\\

Returning to a general discussion, if the Hamiltonian is symmetric under $\pgdel$:
\begin{align}
\pgdel: \;\; \hat{H} = \sum_{\bk} \dg{c}_{\bk,\alpha}\,H(\bk)_{\ab} \pdg{c}_{\bk,\beta}  \;\;\longrightarrow \;\; \hat{H},
\end{align}
then Eq.\ (\ref{blochspatial}) implies
\begin{align} \label{symmonhk}
\hatgdel(\bk)\,H(\boldsymbol{k})\,\hatgdel(\bk)^{\mo} = H\big(\,D_g\boldsymbol{k}\,\big).
\end{align}
By assumption of an insulating gap, $\pdg{\hat{g}}_{\bdelta}(\bk)|u_{n,\bk}\rangle$ belongs in the occupied-band subspace for any occupied band $|u_{n,\bk}\rangle$. This implies a unitary matrix representation (sometimes called the `sewing matrix') of $\pgdel$ in the occupied-band subspace:
\begin{align} \label{goccupied}
 [\calgdel(D_g\bk+\bG,\bk)]_{mn} = \bra{u_{m,D_g\bk+\bG}} \,V(-\bG)\, \pdg{\hat{g}}_{\bdelta}(\bk) \,\ket{u_{n,\bk}}, \ins{with} m,n=1,\ldots,\noc.
\end{align} 
Here, $\bG$  is any reciprocal vector (including zero), and we have applied Eq.\ (\ref{periodP}) which may be rewritten as:
\begin{align} \label{uuk}
\sum_{n =1}^{\noc} \ket{u_{n,\bk}}\bra{u_{n,\bk}} = V(\bG)\,\sum_{n =1}^{\noc} \ket{u_{n,\bk+\bG}}\bra{u_{n,\bk+\bG}} \,V(\bG)^{\mo}.
\end{align}
To motivate Eq.\ (\ref{goccupied}), we are often interested in high-symmetry $\bk$ which are invariant under $\pdg{g}_{\bdelta}$, i.e., $D_g\bk+\bG=\bk$ for some $\bG$ (possibly zero). At these special momenta, the `sewing matrix' is unitarily equivalent to a diagonal matrix, whose diagonal elements are the $\pdg{g}_{\bdelta}$-eigenvalues of the occupied bands. When we're not at these high-symmetry momenta, we will sometimes use the shorthand: $\calgdel(\bk) \equiv \calgdel(D_g\bk,\bk)$, since the second argument is self-evident. We emphasize that $\hatgdel$ and $\brevegdel$ are different matrix representations of the same symmetry element ($\pgdel$), and moreover the matrix dimensions differ: (i) $\hatgdel$ acts on Bloch-combinations of \low orbitals ($\{\phi_{\sma{\boldsymbol{k}, \alpha}}| \alpha=1,\ldots,n_{tot}\}$) defined in Eq.\ (\ref{basisvec}), while (ii) $\brevegdel$ acts on the occupied eigenfunctions ($\{u_{\sma{n,\bk}}|n=1,\ldots, \noc\}$) of $H(\bk)$. \\

It will also be useful to understand the commutative relation between $\hatgdel(\bk)$ and the diagonal matrix $V(\bG)$ which encodes the spatial embedding; as defined in Eq.\ (\ref{aperiodic}), the diagonal elements are $[V(\bG)]_{\ab} = \delta_{\ab} \text{exp}(i\bG \cdot \br_{\alpha})$. We know from Eq.\ (\ref{pgscond2}) that $\boldsymbol{R}^{\sma{g \bdelta}}_{\sma{\ab}}$ is BL vector if $[U_{\sma{g\bdelta}}]_{\alpha \beta} \neq 0$; we further apply the inverse of Eq.\ (\ref{pgscond1}), which states that for any BL vector $\bR$ (which in this context would be $\boldsymbol{R}^{\sma{g \bdelta}}_{\sma{\ab}}$), $D_g^{-1}\bR$ is similarly a BL factor. These two facts combine to give 
\begin{align} 
[U_{\sma{g\bdelta}}]_{\alpha \beta} \neq 0 \imp  D_g^{\mo}\boldsymbol{R}^{\sma{g \bdelta}}_{\sma{\ab}} \in \text{BL} \imp  e^{i\bG\cdot ( \boldsymbol{r_{\beta}}+D_{\sma{g}}^{\mo}\bdelta -D_{\sma{g}}^{\mo}\boldsymbol{r_{\alpha}})}=1,
\end{align}
for a RL vector $\bG$. Applying this equation in 
\begin{align}
0 \neq [\hatgdel(\bk)\, V(\bG)]_{\ab} \eq e^{-i (D_g \boldsymbol{k})\cdot \bdelta }\, [U_{\sma{g\bdelta}}]_{\ab} \,e^{i\bG \cdot \br_{\beta}} = e^{-i (D_g \boldsymbol{k})\cdot \bdelta }\, [U_{\sma{g\bdelta}}]_{\ab} \,e^{i(D_g \bG) \cdot (\br_{\alpha}-\bdelta)} \notag \\
\eq e^{-i(D_g \bG) \cdot \bdelta} \,[V(D_g\bG)\,\hatgdel(\bk)]_{\ab},
\end{align}
we then derive
\begin{align} \label{uv}
\hatgdel(\bk)\, V(\bG) = e^{-i(D_g\bG)\cdot \bdelta}\,V(D_g\bG) \,\hatgdel(\bk),
\end{align}
This equality applies only if the argument of $V$ is a reciprocal vector.

\subsection{Effect of space-time symmetry on the tight-binding Hamiltonian}\label{sec:spacetimetight}

Consider a general space-time transformation $T \pgdel$, where now we include the time-reversal $T$; the following discussion also applies if $\pgdel$ is the trivial transformation.    
\begin{align}
T \pgdel \;: \;\dg{c}_{\alpha}(\boldsymbol{R}+\boldsymbol{r_{\alpha}}) \rightarrow \dg{c}_{\beta}\big(D_g\boldsymbol{R}+\boldsymbol{R}^{\sma{ Tg\bdelta}}_{\sma{\beta \alpha}}+\boldsymbol{r_{\beta}}\,\big) \,[\utg]_{\beta \alpha},
\end{align}
where $\utg$ is the matrix representation of $T \pgdel$ in the \low orbital basis, $\boldsymbol{R}^{\sma{Tg\bdelta}}_{\sma{\beta \alpha}} =D_g\boldsymbol{r_{\alpha}} + \bdelta -\boldsymbol{r_{\beta}}$, 
\begin{align} \label{uitv}
[\utg]_{\beta \alpha} \neq 0 \imp  \boldsymbol{R}^{\sma{Tg\bdelta}}_{\sma{\beta \alpha}} \in \text{BL}, 
\end{align}
and the Bravais-lattice mapping of $\bR$ to $D_g\bR + \boldsymbol{R}^{\sma{Tg\bdelta}}_{\sma{\beta \alpha}}$ is bijective. It follows that the Bloch-wave-transformed \low orbitals transform as
\begin{align}
T \pgdel \;: \;\dg{c}_{\boldsymbol{k},\alpha} \; \longrightarrow \; e^{i (D_g \boldsymbol{k}) \cdot \bdelta } \dg{c}_{-D_g\boldsymbol{k},\beta}\,[\utg]_{\beta \alpha}.
\end{align}
This motivates the following definition for the \low representation of $T \pgdel$:
\begin{align} \label{Tgdrep}
{\hat{T}}_{\sma{g\bdelta}}(\bk) \equiv e^{i (D_g \boldsymbol{k}) \cdot \bdelta } \,\utg\,K,
\end{align}
where $K$ implements complex conjugation, such that a symmetric Hamiltonian ($T \pgdel : \hat{H} \rightarrow \hat{H})$ satisfies
\begin{align} \label{sptiHam}
\hatTg(\bk)\,H(\bk)\,\hatTg(\bk)^{\mo}= H\big({-}D_g\bk\,\big).
\end{align}
For a simple illustration, we return to the lattice of Fig.\ \ref{fig:nonsymmorphic_ex}, where time-reversal symmetry is represented by $\hat{T}(\bk)=-i\sy K$ in a basis where $\sz=+1$ corresponds to spin up in $\vec{z}$. Observe that time reversal commutes with any spatial transformation:
\begin{align}
\text{for} \;\; j \in \{x,z\}, \;\;\;\;\hat{T}(D_j\bk)\;\hat{\bar{M}}_{\sma{j}}(\bk)=\hat{\bar{M}}_{\sma{j}}(-\bk)\;\hat{T}(\bk).
\end{align}
If the Hamiltonian is gapped, there exists an antiunitary representation of $T\pgdel$ in the occupied-band subspace:
\begin{align} \label{sew}
[\caltgdel(\bG-D_g\bk,\bk)]_{mn} = \bra{u_{m,\bG-D_g\bk}} \,V(-\bG)\, {\hat{T}}_{\sma{g\bdelta}}(\bk) \,\ket{u_{n,\bk}}, \ins{where} m,n=1,\ldots,\noc,
\end{align} 
$\bG$ is any reciprocal vector and we have applied Eq.\ (\ref{uuk}). Once again, we introduce the shorthand: $\caltgdel(\bk) \equiv \caltgdel(-D_g\bk,\bk)$. Eq.\ (\ref{uitv}) and (\ref{pgscond1}) further imply that
\begin{align} 
[U_{\sma{Tg\bdelta}}]_{\alpha \beta} \neq 0 \imp  D_g^{\mo}\boldsymbol{R}^{\sma{g \bdelta}}_{\sma{\ab}} \in \text{BL} \imp  e^{i\bG\cdot ( \boldsymbol{r_{\beta}}+D_{\sma{g}}^{\mo}\bdelta -D_{\sma{g}}^{\mo}\boldsymbol{r_{\alpha}})}=1,
\end{align}
which when applied to
\begin{align}
0 \neq [\hatTg(\bk)\, V(\bG)\, K]_{\ab} \eq e^{i (D_g \boldsymbol{k})\cdot \bdelta }\, [U_{\sma{Tg\bdelta}}]_{\ab} \,e^{-i\bG \cdot \br_{\beta}} = e^{i (D_g \boldsymbol{k})\cdot \bdelta }\, [U_{T\sma{g\bdelta}}]_{\ab} \,e^{-i(D_g \bG) \cdot (\br_{\alpha}-\bdelta)} \notag \\
\eq e^{+i(D_g \bG) \cdot \bdelta} \,[V(-D_g\bG)\,\hatTg(\bk)\,K]_{\ab},
\end{align}
leads finally to
\begin{align} \label{UU}
\hatTg(\bk)\,V(\bG) = e^{iD_g\bG\cdot \bdelta}\,V(-D_g\bG)\,\hatTg(\bk).
\end{align}

\section{Case study of $\text{KHgSb}$: symmetry analysis of orbitals, and effective models} \label{app:kp}

\subsection{Band inversion and orbital analysis} \label{app:bandinv}

Our goal is to describe the orbital character of the relevant bands, and also to clarify the band inversion that leads to our topological phase in KHgSb. The first step is to derive a convenient basis that emphasizes the $D^4_{6h}$ crystal symmetries, which we remind the reader include: (i) an inversion ($\cali$) centered around a K ion, which we take as our spatial origin, (ii) the screw rotation $\bar{C}_{6z}$ is a six-fold rotation about $\vec{z}$ followed by a fractional lattice translation ($t(c\vec{z}/2)$), (iii) a screwless three-fold rotation $C_{3z}$, and (iv) the reflections $M_y: (x,y,z) {\rightarrow} (x,{-}y,z)$, $\bar{M}_z{=}t(c\vec{z}/2)M_z$ and $\bar{M}_x {=}t(c\vec{z}/2)M_x$. Our first-principles calculations indicate that bands at the Fermi level are predominantly composed of Hg-6$s$ and Sb-5$p_{x,y,z}$ orbitals. For each atom, we then construct orbitals labelled by $|$atom,orbital,$J_z \rangle$, with $J_z$ the eigenvalue of continuous rotation about $\vec{z}$:
\begin{equation} 
\begin{aligned}
&|Hg,~s_{\frac1{2}},\tfrac{1}{2} \rangle  = \left|is,\uparrow\right\rangle, \myspace &|Hg,~s_{\frac1{2}},-\tfrac{1}{2}\rangle  = \left|is,\downarrow\right\rangle, \\
&|Sb,~p_{\frac{3}{2}},\tfrac{3}{2}\rangle  = \tfrac{1}{\sqrt{2}}\left|-(p_x+ip_y),\uparrow\right\rangle, \myspace &|Sb,~p_{\frac{3}{2}},-\tfrac{3}{2}\rangle   =\tfrac{1}{\sqrt{2}}  \left|(p_x-ip_y),\downarrow\right\rangle, \\
&|Sb,~p_{+},\tfrac{1}{2}\rangle   = \alpha \left|-\tfrac{1}{\sqrt{2}}(p_x+ip_y),\downarrow\right\rangle +\beta\left|p_{z},\uparrow\right\rangle, \myspace &|Sb,~p_{+},-\tfrac{1}{2}\rangle   = \alpha^* \left|\tfrac{1}{\sqrt{2}}(p_x-ip_y),\uparrow\right\rangle + \beta^*\left|p_{z},\downarrow\right\rangle, \\
&|Sb,~p_{-},\tfrac{1}{2}\rangle   = \beta  \left|-\tfrac{1}{\sqrt{2}}(p_x+ip_y),\downarrow\right\rangle-\alpha\left|p_{z},\uparrow\right\rangle, \myspace &|Sb,~p_{-},-\tfrac{1}{2}\rangle   = \beta^* \left|\tfrac{1}{\sqrt{2}}(p_x-ip_y),\uparrow\right\rangle -\alpha^*\left|p_{z},\downarrow\right\rangle,
\end{aligned}
\label{eq:socbasis}
\end{equation}
Here, $\uparrow$ refers to the $+1/2$-eigenstate of spin component $S_z$, and $|\alpha|^2+|\beta|^2=1$ ensures the orthonormality of the $p_{\pm}$ states. Having constructed a set of atomic-centered orbitals, we then take their linear combinations to form eigenstates of inversion, i.e., we form bonding and antibonding states of definite parity:
\begin{align} \label{bonding}
  &|S^{\pm},J_z\rangle=\frac{1}{\sqrt{2}}(|\text{Hg}1,s_{\frac1{2}},J_z\rangle \pm |\text{Hg}2,s_{\frac1{2}},J_z\rangle), \;\;\text{with}\;\; J_z = \pm 1/2 \;\;\text{giving 4 states, and} \notag \\
  &|P_{\alpha}^{\pm},J_z\rangle=\frac{1}{\sqrt{2}}(|\text{Sb}1,p_{\alpha},J_z\rangle \mp |\text{Sb}2,p_{\alpha},J_z\rangle), \;\;\text{with}\;\; \ket{p_{\alpha},J_z} \in \{ \ket{p_{\frac3{2}},\pm \tfrac3{2}},|p_{+},\pm \tfrac{1}{2}\rangle, |p_{-},\pm \tfrac{1}{2}\rangle \} \;\;\text{giving 12 states.}
 \end{align}
Hg1 and Hg2 are illustrated in Fig.\ \ref{fig:eff}(a) as blue atoms which sit diametrically across the inversion center; similarly, Sb1 and Sb2 are red atoms related by spatial inversion; the superscript $\pm$ on $S^{\pm}$ and $P^{\pm}$ indicates an inversion eigenvalue of $\pm 1$. \\

\begin{figure}[H]
\centering
\includegraphics[width=14 cm]{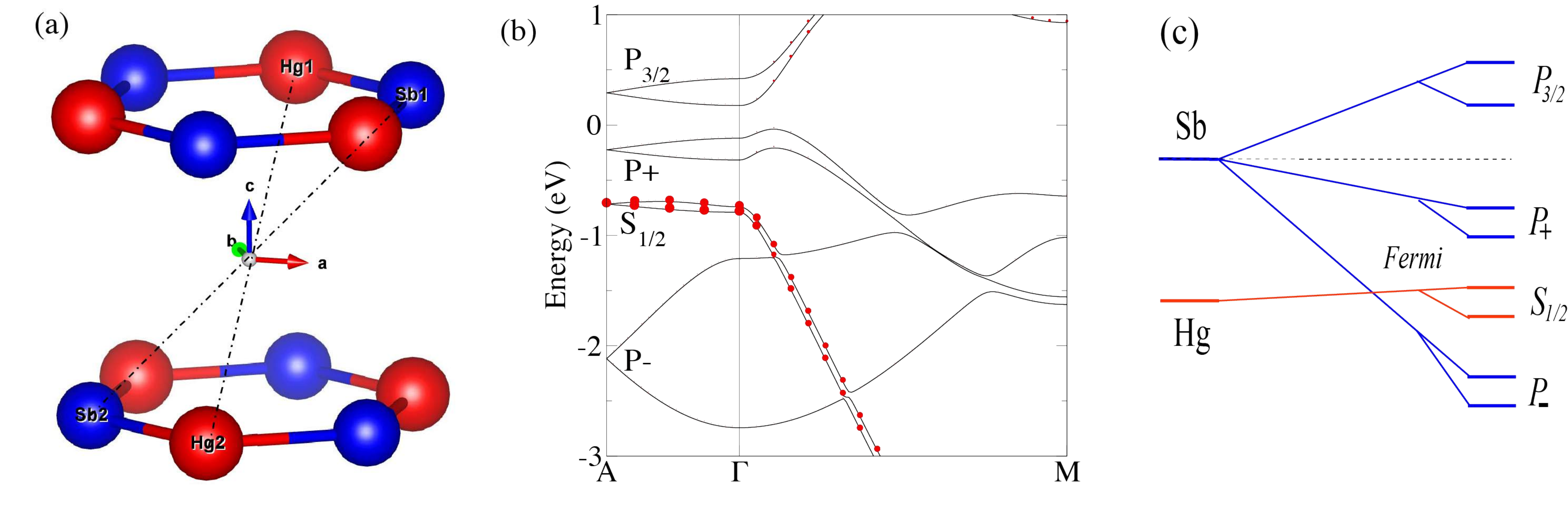}
    \caption{(a) 3D view of atomic structure. The Hg (red) and Sb (blue) ions form a honeycomb layers with AB stacking. Omitted from this picture is the K ion, which is located at the inversion center (also our spatial origin). (b) Each quadruplet is labeled in the band structure. (c) Orbital character of KHgSb at any point along $\Gamma A$, as we vary the crystal field and spin-orbit coupling from zero (leftmost) to their natural strengths (rightmost).  In atomic limit (leftmost), there are only two degenerate levels corresponding respectively to $s$ and $p$ orbitals, due to the emergent SO(3) rotational symmetry.} \label{fig:eff}
\end{figure}

For each of $\kappa \in \{S^{\pm},P^{\pm}_{\alpha}\}$,  $|\kappa,\pm J_z\rangle$ are Kramers-degenerate at time-reversal-invariant momenta (TRIM). The band inversion we will describe occurs at the TRIM points $\Gamma$ and $A {=}(0,0,\pi/c)$; $A$ is special for a further-enhanced four-fold degeneracy, which may be understood in this light: (i) since $A$ also lies on the $k_x{=}0$ glide plane, states divide according to two \emph{real} eigenvalue branches (${\pm}1$) of $\bmx$, (ii) since $[T,\bmx]=0$, Kramers partners have identical $\bmx$-eigenvalues, and (iii) since $\bmx\,{\cal I}=t(c\vec{z})\,{\cal I}\,\bmx$ with $\cali$ the spatial inversion and $t$ a lattice translation, inversion-related partners at $k_z{=}\pi/c$ (and consequently $t(c\vec{z}){=}{-}1$) have opposite $\bmx$-eigenvalues. The net result of (i-iii) is every degenerate subspace at $A$ contains two states in each $\bmx$-subspace, such that one $\bmx$-subspace is related to the other by spatial inversion. Consequently, our basis divides into four quadruplets which are individually connected; we give each quadruplet a name:
\begin{align}
\{|S^\pm,\pm\tfrac{1}{2}\rangle\} \in S_{\frac{1}{2}},\myspace\{|P^\pm_{\frac{3}{2}}, \pm\tfrac{3}{2}\rangle\} \in P_{\frac{3}{2}}, \myspace \{|P^\pm_+, \pm\tfrac{1}{2}\rangle\} \in P_+, \ins{and}\{P^\pm_-, \pm\tfrac{1}{2}\rangle\} \in P_-.
\end{align}
In order of decreasing energy, we have $P_{\frac3{2}}, P_+, S_{\frac1{2}}$ and finally $P_-$, as illustrated in Fig.\ \ref{fig:eff}(b-c). The energy gap at the Fermi level separates $P_{\frac3{2}}$ and $P_+$ quadruplets, and may be attributed to spin-orbit splitting of $p$-type orbitals.  Compared to trivial KZnP (Fig.\ 5(b) in main text), KHgSb has an inverted ordering of $S_{\frac{1}{2}}$ and $P_{\frac{3}{2}}$ quadruplets at both $\Gamma$ and $A$ points; as shown in Fig.\ \ref{fig:eff}(c), this inversion for KHgSb occurs even in the metallic limit of vanishing crystal field and spin-orbit coupling. This band inversion implies a groundstate with mixed Hg-s and Sb-p characters, resulting in a nontrivially-quantized, non-Abelian polarization ($\calq_{\tilg\tilz}$) that we described in the main text; we further demonstrate in App.\ \ref{modelG} that the same inversion also results in a nontrivial mirror Chern number. Since each quadruplet comprises two parity-even  and two parity-odd bands, for KHgSb there is no net parity inversion at any high-symmetry momentum. The classification by time-reversal symmetry is thus trivially (0;000), as is consistent with previous works.\cite{zhang2011,yan2012} Instead, KHgSb and KZnP are distinguished by their rotational quantum numbers, as we explain in App.\ \ref{app:mcnrotate}.\\

We end this section with a further elaboration of origin of the band inversion that distinguishes KHgSB and KZnP. Their difference may be traced back to their atomic limits, where the crystal field and spin-orbit coupling are zero. For KZnP, the atomic $p$ levels (fully-filled) lie below the $s$-levels (unfilled), so that the groundstate is insulating. For KHgSb, the $s$ levels (`fully filled') lie below the $p$-levels (`partially filled'), if we define `filling' by occupying the 12 lowest energy levels of the two non-interacting atoms (Hg and Sb). With this definition of `filling', we might refer to atomic-limit KHgSb as an `atomic topological metal', where the slightest perturbation (e.g., an inter-atomic hopping and/or a spin-orbit coupling) could either (i) lead to the gapped topological phase that is our main subject, or (ii) lead to a Dirac-semimetallic phase in the same equivalence class as Na$_3$Bi,\cite{wang2012} which has the same space group as KHgSb and KZnP; whether scenario (i) or (ii) is selected depends on the relative strengths of the spin-orbit coupling and the crystal field. Further implications of this `atomic topological metal' are left for future investigation.

\subsection{Effective Hamiltonians at the topological phase transition}
\label{symtry}

By comparison of trivial KZnP (Fig.\ 5(b) in main text) with topological KHgSb (Fig.\ \ref{fig:eff}(c)), we have concluded in the last section that both phases differ by a band inversion of the $S_{\frac{1}{2}}$ and $P_{\frac{3}{2}}$ quadruplets. More concretely, we might consider deforming KHgSb (e.g., by pressure/stress) into a trivial phase that is equivalent to KZnP; we might ask what is the effective Hamiltonian that describes this topological phase transition. Since both phases differ only with regard to the filling of the $S_{\frac{1}{2}}$ and $P_{\frac{3}{2}}$ quadruplets, the phase transition can be captured by a minimal model consisting only of these two quadruplets; we will show in App.\ \ref{modelG} that our model accurately describes the change in the mirror Chern number in the $k_z{=}0$ plane, as is alternatively corroborated by a Wilson-loop calculation in App.\ \ref{app:methodswilson}; for completeness, we also show that the mirror Chern number is invariant in the $k_z{=}\pi/c$ plane, which is consistent with a more general symmetry analysis in App.\ \ref{modelA}. The first step to constructing this minimal model is to determine how our symmetries are represented in this reduced basis, as will be done in this section; in later sections we will derive $\bk\cdot \bp$ Hamiltonians that are constrained by these symmetries~\cite{kpbook}. While our $S_{\frac{1}{2}}{-}P_{\frac{3}{2}}$ model is meant to describe changes in topological invariants, we caution that it is not a low-energy description of the actual bulk bands of KHgSb, which near the Fermi level are comprised of $P_{\frac{3}{2}}$ and $P_+$ quadruplets (Fig.\ \ref{fig:eff}(c)).\\

To derive the symmetry representations, it is more convenient to return to the basis of atomic-centered orbitals that we introduced in Eq.\ (\ref{eq:socbasis}), and from which we now extract the relevant orbitals that comprise the $S_{\frac{1}{2}}$ and $P_{\frac{3}{2}}$ quadruplets:
\begin{align} \label{singleatom}
&|\text{Hg}1,s_{\frac1{2}},\frac{1}{2}\rangle,
|\text{Hg}2,s_{\frac1{2}},\frac{1}{2}\rangle,
|\text{Sb}1,p_{\frac{3}{2}},\frac{3}{2}\rangle,
|\text{Sb}2,p_{\frac{3}{2}},\frac{3}{2}\rangle,\notag \\
&|\text{Hg}1,s_{\frac1{2}},-\frac{1}{2}\rangle,
|\text{Hg}2,s_{\frac1{2}},-\frac{1}{2}\rangle,
|\text{Sb}1,p_{\frac{3}{2}},-\frac{3}{2}\rangle, \ins{and}
|\text{Sb}2,p_{\frac{3}{2}},-\frac{3}{2}\rangle.
\end{align}
We define three sets of Pauli matrices: $\sigma_{3}=\pm 1$ corresponds to the sign of $J_z$, $\tau_{3}=1 (-1)$ refers to an $s$ (resp.\ $p$) orbital of Hg (resp.\ Sb), and  $\gamma_3=1(-1)$ to the atomic index $1$ (resp.\ $2$), e.g., $\gamma_1$ flips Hg1 (Sb1) to Hg2 (Sb2). We further define $\sigma_0, \tau_0$ and $\gamma_0$ to be the identity matrix in each corresponding two-dimensional subspace.\\

We apply App.\ \ref{app:reviewtightbind} to derive the symmetry representations in the little group of $\bar{\bk} \in \{\Gamma,A\}$. This group is partially comprised of all spatial transformations ($\pdg{g}_{\bdelta}$) that preserve $\bar{\bk}$ up to a reciprocal vector ($\bG_g(\bar{\bk})$):
\begin{align}
\pdg{g}_{\bdelta}: \;\bar{\bk} \rightarrow D_g\bar{\bk} = \bar{\bk}+\bG_g(\bar{\bk}).
\end{align} 
Following Eq.\ (\ref{actbloch}) and (\ref{act}), its representation in the reduced orbital basis has the general form
\begin{align}
\calr_{\bar{\bk}}(\pdg{g}_{\bdelta}) =V(\bG_g(\bar{\bk})) \,e^{-i (D_g \boldsymbol{k}) \cdot \bdelta }\, U_{\sma{g\bdelta}}.
\end{align}

\noi{1} {\it Spatial inversion} ($\cal{I}$) centered at the K atom  maps atoms as: Hg1 $\leftrightarrow$ Hg2 and Sb1 $\leftrightarrow$ Sb2, as illustrated in Fig.\ \ref{fig:eff}(a); this implies the representation of $\cali$: $\calr_{\bar{\bk}}(\cali)  \propto \gamma_1$. Furthermore, the orbital wavefunction transforms as
\begin{align}
\cali: \myspace \begin{cases} &|s_\frac{1}{2},\pm\frac{1}{2}\rangle \longrightarrow  |s_\frac{1}{2},\pm\frac{1}{2}\rangle, \notag \\
                             &|p_\frac{3}{2},\pm\frac{3}{2}\rangle \longrightarrow -|p_\frac{3}{2},\pm\frac{3}{2}\rangle, \end{cases}
														\end{align}
														implying that $\calr_{\bar{\bk}}(\cali)  \propto \sigma_0 \otimes\tau_3$. In combination,
\begin{align}
\calr_{\bar{\bk}}(\cali)= \sigma_0 \otimes\tau_3\otimes V(-2\bar{\bk})\,\gamma_1, \ins{where} V(-2\bar{\bk})=\left(\begin{array}{cc} e^{-i\bar{k}_z {c}/{2}} &0 \\ 0 & e^{i\bar{k}_z {c}/{2}} \end{array} \right)\end{align}
 for $\bar{k}_z \in \{0,\pi/c\}$.\\

\noi{2} {\it $\bar{C}_{6z}$}, a six-fold rotation about $\vec{z}$ followed  by a fractional lattice translation ($t(c\vec{z}/2)$), maps atoms as: Hg1 $\longleftrightarrow$ Hg2 and Sb1 $\longleftrightarrow$ Sb2. Further applying $\bG_{\sma{C6z}}(\bar{\bk})=0$ and  $e^{-i (D_{\sma{C6z}} \bar{\bk})\cdot \bdelta}=e^{-ik_zc/2}$,
\begin{align}
\calr_{\bar{\bk}}({\bar{C}}_{6z})= e^{-ik_zc/2} [ e^{i\Pi\cdot\pi/3}\otimes \gamma_1],  \ins{with} \Pi=\sigma_3\otimes \left( \begin{array}{cc} 1/2 & 0 \\
    0 & 3/2 \end{array} \right).
		\end{align}

\noi{4} {\it $\bar{C}_{2x}$}, two-fold rotation about $\vec{x}$ followed  by a fractional lattice translation ($t(c\vec{z}/2)$), maps atoms as: Hg1 $\longleftrightarrow$ Hg1 and Sb1 $\longleftrightarrow$ Sb1; it transforms orbitals as
\begin{align}
\bar{C}_{2x}: \myspace \begin{cases} |s_\frac{1}{2},\pm\frac{1}{2}\rangle \longrightarrow -i\,|s_\frac{1}{2},\mp\frac{1}{2}\rangle \notag \\
|p_\frac{3}{2},\pm\frac{3}{2}\rangle \longrightarrow i\,|p_\frac{3}{2},\mp\frac{3}{2}\rangle.\end{cases}
\end{align}
Further applying $e^{-i D_g \boldsymbol{k}\cdot \bdelta}=e^{ik_z\frac{c}{2}}$, we derive
\begin{align}
\calr_{\bar{\bk}}({\bar{C}}_{2x})=e^{ik_zc/2} V(-2\bar{\bk}) [-i\sigma_1\otimes \tau_3 \otimes \gamma_0 ].
\end{align}

\noi{5} {\it $C_{2y}$} maps atoms as Hg1 $\longleftrightarrow$ Hg2 and Sb1 $\longleftrightarrow$ Sb2, and transforms orbitals as 
\begin{align}
\bar{C}_{2y}: \myspace \begin{cases} |s_\frac{1}{2},\pm\frac{1}{2}\rangle \longrightarrow \pm|s_\frac{1}{2},\mp\frac{1}{2}\rangle \notag \\
|p_\frac{3}{2},\pm\frac{3}{2}\rangle \longrightarrow \pm|p_\frac{3}{2},\pm\frac{3}{2}\rangle.\end{cases}
\end{align}
Therefore,
\begin{align} \label{c2y}
\calr_{\bar{\bk}}(\hat{C}_{2y})=-i\,V(-2\bar{\bk})\,\sigma_2\otimes \tau_0 \otimes \gamma_1.
\end{align}

\noi{6} {\it Time reversal} is represented by 
\begin{align}
\calr_{\bar{\bk}}(T) = V(-2\bar{\bk}) \, U_{\sma{T}}\,K =-i\,V(-2\bar{\bk})\,\sigma_2\otimes \tau_0 \otimes \gamma_0 \,K,
\end{align}
where $K$ implements complex conjugation.\\

\noindent These symmetry representations constrain our effective Hamiltonians through Eq.\ (\ref{symmonhk}) and (\ref{sptiHam}), as we now derive.

\subsection {Effective Hamiltonian at $\Gamma$}\label{modelG}

Since our goal is to describe the change in the mirror Chern number in the $k_z{=}0$ plane, we would do well to write our effective Hamiltonian at $\Gamma$ in a basis that diagonalizes the relevant reflection ($\bmz$). At $\Gamma$ where $[\bmz,\cali]=0$, we would also work in a basis that diagonalizes the spatial-inversion operator. Such an inversion eigenbasis has already been found in Eq.\ (\ref{bonding}), from where we obtain the relevant basis vectors: 
 $|S^+,\frac{1}{2}\rangle$,
 $|P^-_{\frac{3}{2}},\frac{3}{2}\rangle$,
 $|S^-,\frac{1}{2}\rangle$,
 $|P^+_{\frac{3}{2}},\frac{3}{2}\rangle$,
 $|S^+,-\frac{1}{2}\rangle$,
 $|P^-_{\frac{3}{2}},-\frac{3}{2}\rangle$,
 $|S^-,-\frac{1}{2}\rangle$,
 $|P^+_{\frac{3}{2}},-\frac{3}{2}\rangle$. To obtain the effective Hamiltonian at $\Gamma$,  we would first transform the symmetry representations derived in the previous section to this inversion eigenbasis, then derive the matrix representation of the Hamiltonian that is consistent with these symmetries. Keeping only the lowest-order terms for each matrix element, the result is
 \begin{eqnarray}
   H_\Gamma(\bf{k}) & = &\left(\begin{array}{cccccccc}
       M_s^+(\bf{k}) & Ak_{+} & 0 &0 & 0& Ck_zk_-^2 &0 & 0 \\
       Ak_{-} & M_p^-(\bf{k}) & 0 &0 & Ck_zk_-^2& 0 &0 & 0 \\
      0 &0 & M_s^-(\bf{k}) & A'k_{+} &0 &0 & 0& C'k_zk_-^2 \\
      0 &0 & A'k_{-} & M_p^+(\bf{k}) &0 &0 & C'k_zk_-^2 &0 \\
      0& Ck_zk_+^2 &0 &0 &  M_s^+(\bf{k}) &-Ak_{-} & 0 &0  \\
      Ck_zk_+^2& 0 &0 &0 & -Ak_{+} & M_p^-(\bf{k}) & 0 &0  \\
    0 &0 & 0& C'k_zk_+^2 & 0 &0 & M_s^-(\bf{k}) &-A'k_{-}  \\
    0 &0 & C'k_zk_+^2 &0 & 0 &0 &-A'k_{+} & M_p^+(\bf{k})  \\
 \end{array}\right)
 \end{eqnarray}
where $k_{\pm}=k_{x}\pm ik_{y}$,
\begin{align} \label{defineM}
M_\alpha^\pm({\bf k})=-M^\pm_{\alpha0} + M^\pm_{\alpha1}k_{z}^{2}+M^\pm_{\alpha2}(k_x^{2}+k_y^2),\\
 \label{definedelta} M^\pm_{\alpha\beta}=M_{\alpha\beta}\pm\Delta_{\alpha\beta}; \;\;\;\; \alpha\in \{s,p\},\;\beta \in \{0, 1,2\},
\end{align}
and $M_{\alpha\beta}$ and $\Delta_{\alpha\beta}$ are parameters fitted to our ab-initio
calculation, as shown in Tab.~\ref{params}.  Where $k_z=0$, this Hamiltonian diagonalizes into two four-by-four blocks, which we distinguish by $\bmz=\pm i$; each block describes two massive Dirac fermions of the same chirality. KHgSb is described by the inverted masses: $M_{s0}^\pm > M_{p0}^\mp$, resulting in a Chern number $\calc_e=2$ in the $\bmz=+i$ subspace; note that the Chern number in the $\bmz=-i$ subspace equals to $-2$, as required by time-reversal symmetry.

\begin{table}[htb]
\begin{center}
\begin{tabular}{c|c|c|c|c|c|c}
  \hline
  \hline
 &$M_{s0}\ (eV)$& $M_{s1}\ (eV\ang^2)$& $M_{s2}\ (eV\ang^2)$& $M_{p0}\ (eV\ang^2)$& $M_{p1}\ (eV\ang^2)$&  $M_{p2}\ (eV\ang^2)$ \\
\hline
 $\Gamma$ & 0.2181  &-0.1 & 49.0000 & -0.2985&-0.1 &-10.00  \\
 $A$      & 0.2181  &-0.1 & 49.0000 & -0.2985&-0.1 &-10.00  \\
\hline
\hline
 &$\Delta_{s0}$ & $\Delta_{s1}  $ & $\Delta_{s2}    $& $\Delta_{p0}  $ & $\Delta_{p1}   $&  $\Delta_{p2}  $  \\
\hline
 $\Gamma$ &-0.0988 & -1.0464 & -1.0000 & 0.1218 & 1.2906 & 0.05  \\
\hline
\hline
     & A $(eV\ang)$ & $A'\ (eV\ang)$ & B $(eV\ang)$ &$B'\ (eV\ang)$& D  $(eV\ang^2)$& F  $(eV\ang^2)$   \\
\hline
 $\Gamma$ & 3.6 & 3.4&$\varnothing$ &$\varnothing$ &$\varnothing$ &$\varnothing$   \\
 $A$ & 3.5 &$\varnothing$ &0.493    &0.523 & 1.0 & 2.0  \\
\hline
\hline
\end{tabular}
\caption{\label{paramts}The fitted parameters for the 8-band model at both $\Gamma$ and $A$ points.}
\label{params}
\end{center}
\end{table}

\subsection {Effective Hamiltonian at $A$} \label{modelA}

To determine the $\bmz$-Chern number in the $k_z=\pi/c$ plane, we would like our effective Hamiltonian at $A$ to be written in an eigenbasis of $\bmz$. Unlike at $\Gamma$, the representations of $\bmz$ and $\cali$ now anticommute at $A$, and so we would not use the inversion eigenbasis that we applied last section. The $\bmz$ eigenbasis is actually comprised of the atomic-centered orbitals of Eq.~(\ref{singleatom}), which we reorder as: 
\begin{align} 
&|\text{Hg}1,s_{\frac1{2}},\frac{1}{2}\rangle,
|\text{Sb}1,p_{\frac{3}{2}},\frac{3}{2}\rangle,
|\text{Hg}2,s_{\frac1{2}},-\frac{1}{2}\rangle,
|\text{Sb}2,p_{\frac{3}{2}},-\frac{3}{2}\rangle,\notag \\
&|\text{Hg}1,s_{\frac1{2}},-\frac{1}{2}\rangle,
|\text{Sb}1,p_{\frac{3}{2}},-\frac{3}{2}\rangle,
|\text{Hg}2,s_{\frac1{2}},\frac{1}{2}\rangle, \ins{and}
|\text{Sb}2,p_{\frac{3}{2}},\frac{3}{2}\rangle,
\end{align}
so that the first (second) four states transform in the ${+}i$ (resp.\ ${-}i$) representation of $\bmz$. Keeping only the lowest-order terms for each matrix element, the effective Hamiltonian at $A$ reads as:
\begin{eqnarray}
H_A(\tilde{\bf{k}})  = &
\left(
\begin{array}{cccccccc}
 M_s(\tilde{\bf{k}}) & A \tilde{k}_+ & 0 & i F \tilde{k}_-^2 & 0 & C \tilde{k}_-^2 \tilde{k}_z & -i B \tilde{k}_z & -i D \tilde{k}_+ \tilde{k}_z \\
 A \tilde{k}_- & M_p(\tilde{\bf{k}}) & i F \tilde{k}_-^2 & 0 & C \tilde{k}_-^2 \tilde{k}_z & 0 & -i D \tilde{k}_- \tilde{k}_z & -i B' \tilde{k}_z \\
 0 & -i F \tilde{k}_+^2 & M_s(\tilde{\bf{k}}) & -A \tilde{k}_- & i B \tilde{k}_z & -i D \tilde{k}_- \tilde{k}_z & 0 & C \tilde{k}_+^2 \tilde{k}_z \\
 -i F \tilde{k}_+^2 & 0 & -A \tilde{k}_+ & M_p(\tilde{\bf{k}}) & -i D \tilde{k}_+ \tilde{k}_z & i B' \tilde{k}_z & C \tilde{k}_+^2 \tilde{k}_z & 0 \\
 0 & C \tilde{k}_+^2 \tilde{k}_z & -i B \tilde{k}_z & i D \tilde{k}_- \tilde{k}_z & M_s(\tilde{\bf{k}}) & -A \tilde{k}_- & 0 & i F \tilde{k}_+^2 \\
 C \tilde{k}_+^2 \tilde{k}_z & 0 & i D \tilde{k}_+ \tilde{k}_z & -i B' \tilde{k}_z & -A \tilde{k}_+ & M_p(\tilde{\bf{k}}) & i F \tilde{k}_+^2 & 0 \\
 i B \tilde{k}_z & i D \tilde{k}_+ \tilde{k}_z & 0 & C \tilde{k}_-^2 \tilde{k}_z & 0 & -i F \tilde{k}_-^2 & M_s(\tilde{\bf{k}}) & A \tilde{k}_+ \\
 i D \tilde{k}_- \tilde{k}_z & i B' \tilde{k}_z & C \tilde{k}_-^2 \tilde{k}_z & 0 & -i F \tilde{k}_-^2 & 0 & A \tilde{k}_- & M_p(\tilde{\bf{k}}) \\
\end{array}
\right) 
\end{eqnarray}
where $ \tilde{{\bf k}}={\bf k}-(0,0,\pi/c)$, $\tilde{k}_{\pm} = \tilde{k}_x \pm i \tilde{k}_y$, and 
$ M_\alpha(\tilde{\bk})=-M_{\alpha0} + M_{\alpha1}\tilde{k}_{z}^{2}+M_{\alpha2}(\tilde{k}_x^{2}+\tilde{k}_y^2)$ for $\alpha \in \{s,p\}$; the fitted parameters $M_{\ab}$ are found in Tab.\ \ref{paramts}. Within the mirror-invariant plane ($k_z=\pi/c$), this Hamiltonian is block-diagonalized as
\begin{eqnarray}
 H_{k_z=\pi/c}(\tilde{k}_x,\tilde{k}_y) =H_A({\tilde{\bf k}}) & \;\stackrel{\mathclap{\normalfont\mbox{$\sma{\tilde{k}_z=0}$}}}{=}\; &\left(\begin{array}{cccccccc}
     M_s(\bf{\tilde{k}}) & A\tilde{k}_{+} & 0 & iF\tilde{k}_-^2  & 0& 0 &0 & 0 \\
     A\tilde{k}_{-} & M_p(\bf{\tilde{k}}) & iF\tilde{k}_-^2 & 0  & 0& 0 &0 & 0 \\
     0 & -iF\tilde{k}_+^2 & M_s(\bf{\tilde{k}}) &-A\tilde{k}_{-} &0 &0 & 0& 0 \\
     -iF\tilde{k}_+^2 & 0 & -A\tilde{k}_{+} & M_p(\bf{\tilde{k}}) &0 &0 & 0&0 \\
   0& 0 &0 & 0 &  M_s(\bf{\tilde{k}}) &-A\tilde{k}_{-} & 0 & iF\tilde{k}_+^2   \\
   0& 0 &0 & 0 & -A\tilde{k}_{+} & M_p(\bf{\tilde{k}}) & iF\tilde{k}_+^2 & 0   \\
  0 &0 & 0& 0 & 0 &-iF\tilde{k}_-^2   & M_s(\bf{\tilde{k}}) &A\tilde{k}_{+}  \\
  0 &0 & 0& 0 &-iF\tilde{k}_-^2 & 0   &A\tilde{k}_{-} & M_p(\bf{\tilde{k}})  \\
 \end{array}\right);
\end{eqnarray} 
 each block, corresponding to one mirror representation, comprises two massive Dirac fermions with opposite chirality. Both masses are identical by $C_{2y}$ symmetry, as follows from: (i) $\C_{2y}$ representatively commuting with $\bmz$ at A, and (ii) $C_{2y}$ mapping Hg1 $\longleftrightarrow$ Hg2 and Sb1 $\longleftrightarrow$ Sb2, as recalled from Eq.\ (\ref{c2y}). The identity of both masses ensures there is never any net chirality, i.e., the mirror Chern number vanishes in the $k_z{=}\pi/c$ plane that contains $A$; this vanishing is a general feature of this space group, as we alternatively prove in App.\ \ref{app:kzpi}.

\section{Diagnosing bulk topological invariants}\label{app:bulkinvariants}

Our aim is to diagnose topological invariants from bulk wavefunctions. As described in the main text, one useful diagnosis tool is the spectrum of the projected-position operator ($\pper\hat{y}\pper$). Since the position operator ($\hat{y}$) commutes with translations parallel to the surface, we would define 
\begin{align}
\pper(\kpar) = \sum_{n=1}^{\noc}\int_{-2\pi/a}^{2\pi/a} \frac{dk_y}{(4\pi/a)} \,\ketbra{\psi_{n,k_y,\kpar}}{\psi_{n,k_y,\kpar}}
\end{align}
to project to all occupied bands with surface wavevector $\kpar$, and $\psi_{n,\bk}(\br)=$exp$(i\bk\cdot \br)u_{n,\bk}(\br)$ are the Hamiltonian eigenstates in Bloch-wave form; here we have chosen an unconventional ordering for the bulk wavevector: $\bk \equiv (k_y,k_x,k_z) \equiv (k_y,\kpar)$, and $a/2 {=}\tilde{\boldsymbol{a}}_1{\cdot} \vec{y}$ with $\tilde{\boldsymbol{a}}_1$ equal to the lattice vector indicated in Fig.\ \ref{fig:wherewilson}(a). App.\ \ref{sec:connectWilson} reveals the easiest way to calculate the spectrum of $\pper\hat{y}\pper$ -- by a well-known relation between $\pper\hat{y}\pper$ and the Wilson loop,\cite{AA2014} which is the matrix representation of holonomy. Building upon this, we then propose an efficient way to extract the mirror Chern number in crystals with time-reversal symmetry, and as a case study, we applied our method to the $k_z=0$ mirror plane of KHgSb in App.\ \ref{app:methodswilson}. In contrast, the mirror Chern number in the $k_z=\pi/c$ plane must vanish for our space group, as we demonstrate in App.\ \ref{app:kzpi}. Finally in App.\ \ref{app:noQSHE}, we argue that there can be no quantum spin Hall effect in the $k_x=\pi/\sqrt{3}a$ glide plane.

\begin{figure}[H]
\centering
\includegraphics[width=13 cm]{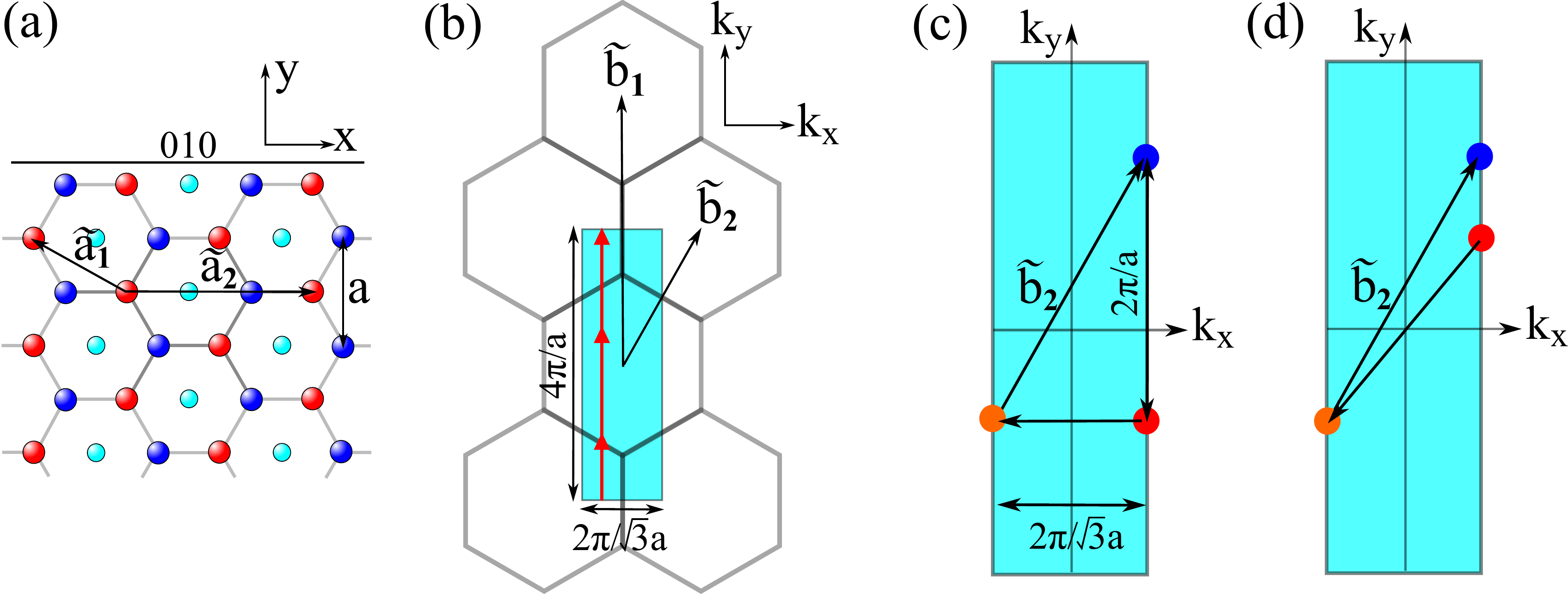}
    \caption{(a) Top-down view of atomic structure, with two of three Bravais lattice vectors indicated by $\tilde{\boldsymbol{a}}_1$ and $\tilde{\boldsymbol{a}}_2$. (b) A constant-$k_z$ slice of the space of crystal momentum, with two of three reciprocal lattice vectors indicated by $\tilde{\boldsymbol{b}}_1$ and $\tilde{\boldsymbol{b}}_2$. While each hexagon corresponds to a Wigner-Seitz primitive cell, it is convenient for this Section to pick the rectangular primitive cell that is shaded in cyan. (c) is a close-up of the rectangular primitive cell in (b). Here in (c), we illustrate how the glide reflection ($\bmx$) maps $(k_y,\pi/\sqrt{3}a,k_z) \rightarrow (k_y,-\pi/\sqrt{3}a,k_z)$ (red dot to brown)  which connects to $(2\pi/a+k_y,\pi/\sqrt{3}a,k_z)$ (blue) through $\tilde{\boldsymbol{b}}_2$. The same mapping by $\bmx$ in the hexagonal primitive cell can be viewed in (e). Figure (d) should be interpreted as the $k_z=0$ cross-section, and illustrates the effect of time-reversal, which maps  $(k_y,\pi/\sqrt{3}a,0) \rightarrow (-k_y,-\pi/\sqrt{3}a,0)$ (red dot to brown),  which then connects to $(2\pi/a-k_y,\pi/\sqrt{3}a,0)$ (blue) through $\tilde{\boldsymbol{b}}_2$.} \label{fig:wherewilson}
\end{figure}

\subsection{Review of Wilson loops and their connection to the projected-position operator} \label{sec:connectWilson}

The spectrum of the projected-position operator ($\pper(\kpar)\hat{y}\pper(\kpar)$) is obtained by diagonalizing a Wilson-loop operator, which effects parallel transport of the occupied bands along a non-contractible loops in the BZ. We consider a family of loops parametrized by $\kpar =(k_x \in [-\pi/\sqrt{3}a,+\pi/\sqrt{3}a],k_z\in [-\pi/c,+\pi/c])$, where for each loop $\kpar$ is fixed while $k_y$ is varied over a non-contractible circle (colored red in Fig.\ \ref{fig:wherewilson}(b)). In the \low-orbital basis, such transport is represented by the Wilson-loop operator
\begin{align} 
\hat{W}(\kpar) = V(4\pi \vec{y}/a)\prod_{k_y}^{2\pi/a \leftarrow -2\pi/a} P(k_y,\kpar),
\end{align}
where we have discretized the momentum as $k_y = 4\pi m/(a N_y)$ for integer $m=1,\ldots,N_y$ and $4\pi \vec{y}/a$ a reciprocal vector, and $(2\pi/a{\leftarrow}{-}2\pi/a)$ indicates that the product of projections is path-ordered in the direction of increasing $k_y$. The role of the path-ordered product is to map a state in the occupied subspace ($\calh(-2\pi/a,\kpar)$) at $(-2\pi/a,\kpar)$ to one ($|{\tilde{u}}\rangle$) in the occupied subspace at $(2\pi/a,\kpar)$; the effect of $V(4\pi \vec{y}/a)$ is to subsequently map $|{\tilde{u}}\rangle$ back to $\calh(-2\pi/a,\kpar)$, thus closing the parameter loop; cf.\ Eq.\ (\ref{aperiodic}). In the limit of large $N_{\sma{y}}$, $\noc$ eigenvalues of $\hat{W}$ become unimodular, and we label them by exp$[i\theta_{\sma{n,\kpar}}]$ with $n=1,\ldots, \noc$. Denoting the eigenvalues of $\pper(\kpar)\hat{y}\pper(\kpar)$ as ${y_{\sma{n,\kpar}}}$, the two spectra are related as $y_{\sma{n,\kpar}}/(a/2) = \theta_{\sma{n,\kpar}}/2\pi$ modulo one.\cite{AA2014}

\subsection{Mirror Chern number in the $k_z=0$ mirror plane} \label{app:methodswilson}

For time-reversal-invariant crystals, we propose an efficient method to calculate the mirror Chern number ($\calc_e$) through the Wilson loop -- by exploiting the time-reversal symmetry, we are able to extract $\calc_e$ from wavefunctions in \emph{half} of a mirror plane. \\

Our topological invariant is defined as the integral of the Berry curvature\cite{berry1984} ($\calf_e$) over the $k_z=0$ mirror plane, as contributed by the even ($\bmz=+i$) subspace of reflection:
\begin{align}
\calc_e = \frac1{2\pi}\int_0^{2\pi} dk_x \int_0^{2\pi} dk_y \,\calf_e(k_x,k_y).
\end{align} 
Recall here that $\bmz$ is a normal reflection that squares to a $2\pi$ rotation, hence its representation has eigenvalues $\pm i$. In spin-orbit-coupled systems, time-reversal symmetry relates even and odd ($M_z=-i$) subspaces by $\calf_e(k_x,k_y) = -\calf_o(-k_x,-k_y)$, and therefore we re-express
\begin{align} \label{mcnre}
\calc_e = \frac1{2\pi}\int_0^{\pi} dk_x \int_0^{2\pi} dk_y \,(\,\calf_e(k_x,k_y)-\calf_o(k_x,k_y)\,)
\end{align} 
as an integral over half of a mirror plane. By Stoke's theorem, we may relate an integral of curvature to differences in the Wilson-loop phases ($\{\theta\}$) between $k_x{=}0$ and $\pi$.\cite{kingsmith1993,AA2014} Due to the orthogonality of the mirror subspaces, we may label each $\theta$-band by its mirror eigenvalue: $\theta^e$ ($\theta^o$) in the even (odd) subspace is colored red (blue) in Fig.\ \ref{fig:wilson_mcn}. $\calc_e$ is thus further rewritten as the net change in $\theta^o$ in the interval $k_x \in [0,\pi]$, minus the net change in $\theta^e$:
\begin{align} \label{winding}
\calc_e = \frac{1}{2\pi}\,\int_0^{\pi} dk_x\,\sum_{i=1}^{\noc/2}\bigg(\,\partdif{\theta^o_i}{k_x} - \partdif{\theta^e_i}{k_x}\,\bigg).
\end{align}
Here, we have distinguished different Wilson-loop phases by a band index in the subscript of $\theta$; given $\noc$ occupied bands, time-reversal symmetry ensures an even split between even and odd representations, which we label respectively by $\theta^e_i$ and $\theta^o_i$ with $i{=}1,\ldots,\noc/2$. $\calc_e$ is most easily extracted from $\{\theta\}$ by a single-phase criterion: consider the intersections of $\{\theta\}$ with an arbitrary constant-phase line. At each intersection, we evaluate [sign of $\partial\theta/\partial k_x$] $\times$ [mirror eigenvalue$/i$], then sum this quantity over all intersections along $k_x {\in} [0,\pi]$. Applying this method to Fig.\ \ref{fig:wilson_mcn}, we find $\calc_e=2$ for KHgSb. This result is further supported by our analysis of the rotational eigenvalues in App.\ \ref{app:mcnrotate}. \\

\begin{figure}[H]
\centering
\includegraphics[width=8 cm]{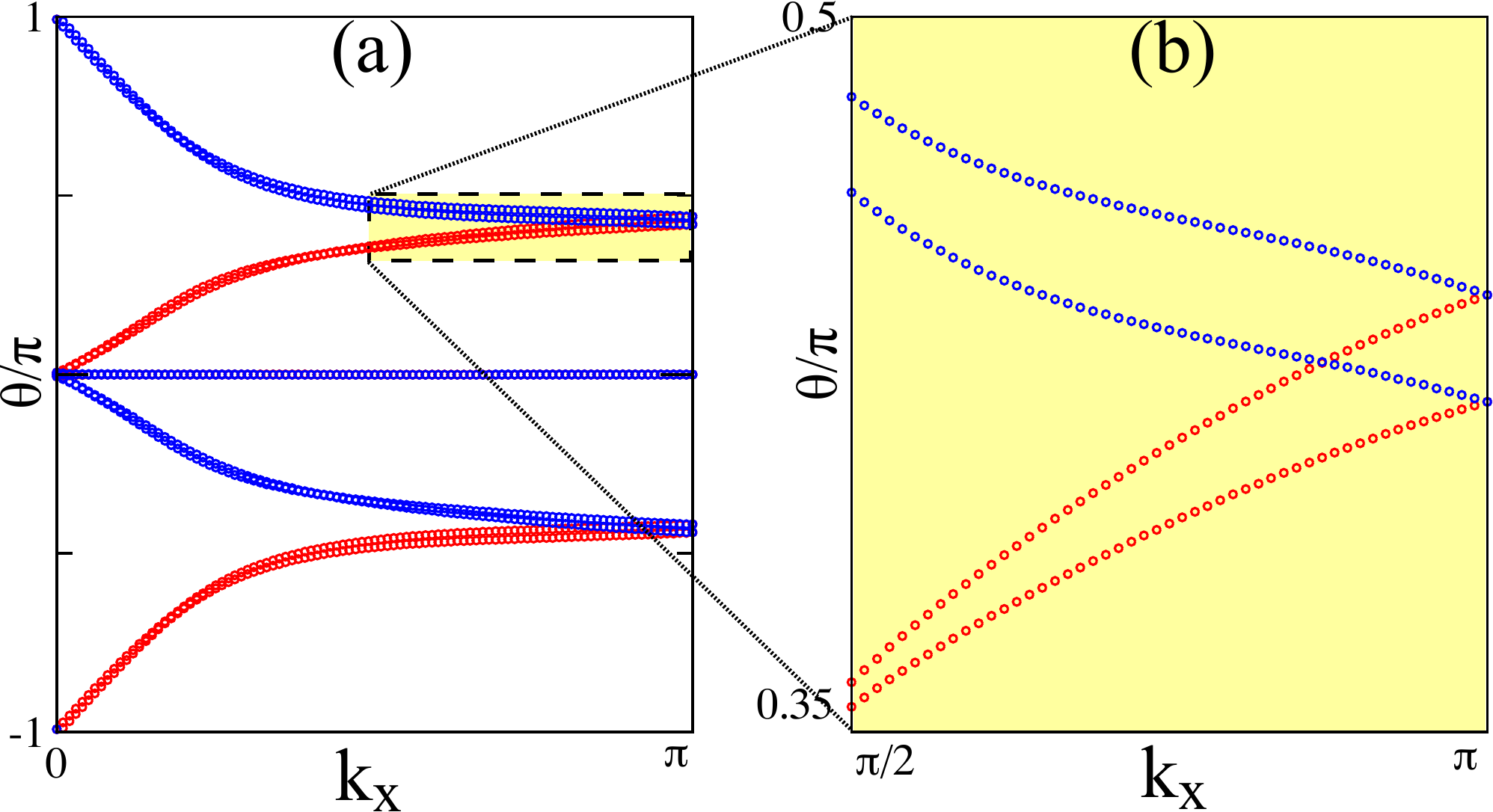}
    \caption{(a) Wilson-loop spectrum along $\tilde{\Gamma}\tilde{Z}$. The mirror eigenvalue of each band is indicated by color (red for $+i$ and blue for $-i$) and also directly by $\pm i$ labels in the figure. (b) is a close-up of (a). } \label{fig:wilson_mcn}
\end{figure}

For integer-spin systems with a mirror symmetry ($M$) satisfying $M^2=I$, we remark that time reversal instead relates  $\calf_{e}(k_x,k_y) = -\calf_{e}(-k_x,-k_y)$ and $\calf_{o}(k_x,k_y) = -\calf_{o}(-k_x,-k_y)$, where even and odd representations now correspond to $M=+1$ and $-1$ subspaces, respectively. Consequently, the mirror Chern number $\calc_e$, also defined as the integral of $\calf_e$ over the mirror plane, vanishes. Nevertheless, the right-hand side of Eq.\ (\ref{mcnre}) is independently valid as a different topological invariant, which we have shown to be quantized in rotationally-symmetric crystals.\cite{AAchen}

\subsection{Vanishing of mirror Chern number in the $k_z=\pi/c$ plane} \label{app:kzpi}

Given that the rotational inversion along $\Gamma A$ leads to nontrivial band topology in the $k_z=0$ plane, we might ask if the $k_z=\pi/c$ mirror plane manifests the same topology. We find for the latter plane that spin-degenerate partners (related by the space-time inversion $T\cali$) belong in the same $\bmz$ subspace, thus contributing canceling Berry curvatures at each momentum; in contrast, spin-degenerate partners in the $k_z=0$ plane belong in opposite mirror subspaces. This is more generally true for any space group with inversion and glideless-mirror symmetries that preserve different origins, i.e., $\bmz \cali = t(c\vec{z})\,\cali \bmz$.\\

\noindent \textbf{Proof} At each wavevector ($\tilde{\boldsymbol{k}}$) in this mirror plane, the mirror-projected Berry curvatures must vanish, i.e., $\calf_e(\bk)=\calf_o(\bk)=0$, as we now demonstrate. The group of $\tilde{\boldsymbol{k}}$  comprises  $T\cali$ and $\bmz$, whose representations anticommute due to two reasons: (i) From $\bmz \cali = t(c\vec{z})\,\cali \bmz$, the translation ($t$) acts on a Bloch wave to produce a phase factor (exp$[-{ik_zc}])$ which equals $-1$ since $k_z=\pi/c$. (ii) Being spatially local, time reversal commutes with any space group element. It follows that if $|\psi(\tilde{\boldsymbol{k}})\rangle$ is an eigenstate of $\bmz$ with eigenvalue $\pm i$, its spin-degenerate partner ($\,T\cali |\psi(\tilde{\boldsymbol{k}})\rangle \,$) belongs in the same mirror subspace. Since $T \cali$ is antiunitary, spin-degenerate partners contribute canceling Berry curvatures, thus ruling out a quantum anomalous Hall effect {within} the same mirror subspace.

\subsection{No QSHE in the $k_x= \pi/\sqrt{3}a$ glide plane}\label{app:noQSHE}

We remind the reader through Fig.\ \ref{fig:wherewilson}(b) that the reciprocal vectors are
\begin{align} \label{reciprocalvectors}
\tilde{\bb}_1 = \tfrac{4\pi}{a}\vec{y},\;\; \tilde{\bb}_2 = \tfrac{2\pi}{\sqrt{3}a}\vec{x} + \tfrac{2\pi}{a}\vec{y}, \ins{and} \tilde{\bb}_3 = \tfrac{2\pi}{c}\vec{z}.
\end{align}
Fig.\ \ref{fig:wherewilson}(c-e) further illustrate that the $k_x= \pi/\sqrt{3}a$ plane is invariant under both glide-reflection and time-reversal symmetries. In the first step, we formulate a QSH topology in this plane by assuming \emph{only} time-reversal symmetry, and then we show the effect of glide symmetry is to rule out the QSH phase altogether.  \\

The time-reversal-invariant momenta in this glide plane lie at $(k_y,k_z) \in \{(\pm\pi/a,0),(\pm\pi/a,\pi/c)\}$, as follows from time reversal mapping $\bk = (\pi/\sqrt{3}a,k_y,k_z) \rightarrow -\bk$,  which further connects to $(\pi/\sqrt{3}a,2\pi/a-k_y,-k_z)$ through the reciprocal vector $\tilde{\bb}_2$, as illustrated in Fig.\ \ref{fig:wherewilson}(d). We follow the Kane-Mele formulation\cite{kane2005B} of the $\Z_2$ invariant by first defining the matrix
\begin{align} \label{skewsew}
[\cala_{\sma{\bk}}]_{ij} = \bra{u_{\sma{i,\bk}}}\,\hat{T}\,\ket{u_{\sma{j,\bk}}}\,K, \ins{with} i,j =1,\ldots, \noc,
\end{align}
where $K$ implements complex conjugation and time reversal is represented by the anti-unitary operator $\hat{T}=U_{\sma{T}}K$. $\hat{T}^2{=}{-}I$ then implies $\cala_{\sma{\bk}}$ is skew-symmetric, so we may define its Pfaffian by $\zeta_{\sma{\bk}}= \text{Pf}[\cala_{\sma{\bk}}]$. The Kane-Mele criterion for a QSH phase is an odd number of zeros of $\zeta$ in half the glide plane,\cite{kane2005B} which implies at least one of these zeros can never be annihilated, e.g., see Fig.\ \ref{fig:pfaffianzeros}(a). To be concrete, we take the half-glide plane with $k_z \in [0,\pi/c]$. To simplify our argument, we have assumed the zeros of $\zeta$ form isolated points instead of lines; it is known with spatial-inversion symmetry that $\zeta$ is real and its zeros, if any, form lines in the plane.\cite{kane2005B} Supposing a quantum spin Hall insulator were also inversion symmetric, it would remain in the same topological phase (as classified by the time-reversal-invariant $\Z_2$ index\cite{kane2005B}) if inversion symmetry is softly broken, while preserving both the energy gap and the time reversal symmetry. Since the goal of this section is to identify and eventually rule out the QSH phase, we simply assume that spatial-inversion symmetry is absent. To rule out the QSH phase in the presence of glide-mirror symmetry, we now demonstrate that zeros of $\zeta$, if they exist in isolated points, can always mutually annihilate.\\

\begin{figure}[ht]
\centering
\includegraphics[width=8 cm]{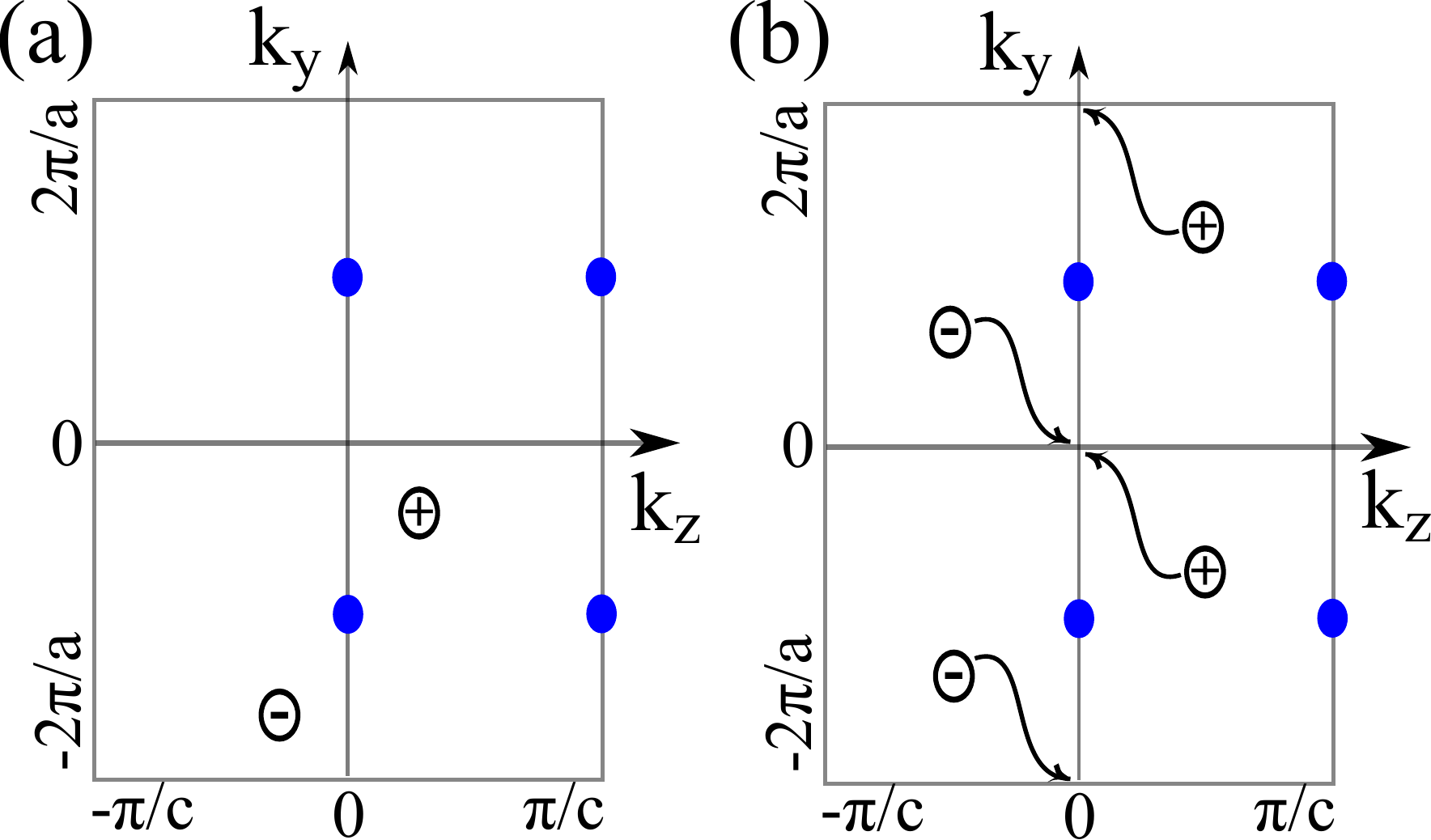}
    \caption{Zeros of the Pfaffian are indicated by empty circles, and the vorticity of each zero is indicated by the $\pm$ sign within each circle; time-reversal-invariant momenta are indicated by solid blue dots. (a) Hypothetical zeros in a quantum spin Hall phase with \emph{only} time-reversal symmetry. (b) In addition with glide symmetry, zeros minimally come in quadruplets which can always mutually annihilate; the curved arrows show an example of zero trajectories which lead to annihilation.} \label{fig:pfaffianzeros}
\end{figure}

\noindent {\textbf{Proof}} Consider the glide reflection $\bmx$, which transforms spatial coordinates as $(x,y,z)\rightarrow (-x,y,z+1/2)$. In the glide plane, $\bmx$ maps between two momenta which are separated by half a reciprocal period ($\tilde{\bb}_1/2$): 
\begin{align}
\bmx:\myspace (k_y,\pi/\sqrt{3}a,k_z) \longrightarrow (k_y,-\pi/\sqrt{3}a,k_z) = (k_y+2\pi/a,\pi/\sqrt{3}a,k_z)-\tilde{\bb}_2 =(k_y,\pi/\sqrt{3}a,k_z)+\tilde{\bb}_1/2-\tilde{\bb}_2,
\end{align}
as illustrated in Fig.\ \ref{fig:wherewilson}(c); recall here that the reciprocal vectors $\tilde{\bb}_j$ are defined in Eq.\ (\ref{reciprocalvectors}). The Bloch eigenfunctions at $(k_y,\kpar)$ and $(k_y+2\pi/a,\kpar)$ are therefore related through
\begin{align} \label{hooboy}
\ket{u_{\sma{m,k_y+2\pi/a}}} = e^{-ik_zc/2}\sum_{n=1}^{\noc}[\bar{\calu}_{\sma{k_y +2\pi/a\leftarrow k_y}}]_{mn}^* \,V({-}\tilde{\boldsymbol{b}}_2)\,\ubmx\,\ket{u_{\sma{n,k_y}}},
\end{align}
where exp$(-ik_zc/2)\,\ubmx$ represents $\bmx$ in the Bloch-orbital basis, $\bar{\calu}$ is a unitary matrix that `sews' together occupied bands at $(k_y,\kpar)$ and $(k_y+2\pi/a,\kpar)$. Here and henceforth, $\kpar {\equiv} (k_x,k_z)$ is a constant parameter that will be suppressed. Combining this glide constraint with Eq.\ (\ref{skewsew}),
\begin{align} \label{interr}
&[\cala_{\sma{k_y+2\pi/a}}]_{ij} =\sum_{\alpha,\beta=1}^{n_{tot}}u_{\sma{i,k_y+2\pi/a}}(\alpha)^*\,[\ut]_{\ab}\,K\,u_{\sma{j,k_y+2\pi/a}}(\beta)\,K\notag \\ 
\eq \sum_{\alpha,\beta=1}^{n_{tot}}u_{\sma{i,k_y+2\pi/a}}(\alpha)^*\,[\ut]_{\ab}\,u_{\sma{j,k_y+2\pi/a}}(\beta)^*\notag \\
\eq e^{ik_zc}\sum_{\mu,\nu=1}^{n_{tot}}\sum_{m,n=1}^{\noc} [\bar{\calu}_{\sma{k_y +2\pi/a\leftarrow k_y}}]_{im}\,u_{\sma{m,k_y}}(\mu)^*\,\big[\;\dg{\ubmx}\,V(\tilde{\boldsymbol{b}}_2)\,\ut\,V(\tilde{\boldsymbol{b}}_2)\,\ubmx^*\;\big]_{\mu \nu}\,u_{\sma{n,k_y}}(\nu)^*\,[\bar{\calu}^t_{\sma{k_y +2\pi/a\leftarrow k_y}}]_{nj}.
\end{align}
Two more identities are useful: (i) since time reversal commutes with spatial transformations, $\ut \ubmx^* = \ubmx \ut$, and (ii) $\ut \,V(\tilde{\boldsymbol{b}}_2) = V(-\tilde{\boldsymbol{b}}_2)\,\ut$ follows from identifying $D_{\sma{g}}=I$ and $\bdelta=0$ in Eq.\ (\ref{UU}). Combining these identities with Eq.\ (\ref{interr}), we are led to
\begin{align}
\cala_{\sma{k_y+2\pi/a}} = e^{ik_zc}\,\bar{\calu}_{\sma{k_y +2\pi/a\leftarrow k_y}}\,  \cala_{\sma{k_y}}\,\bar{\calu}_{\sma{k_y +2\pi/a\leftarrow k_y}}^t. 
\end{align}
Applying a well-known Pfaffian identity, we conclude that 
\begin{align} \label{pfaffianresult}
\zeta_{\sma{k_y+2\pi/a,\kpar}} = e^{i\noc k_zc/2}\,\text{det}[\,\bar{\calu}_{\sma{k_y +2\pi/a\leftarrow k_y}}\,]\;\zeta_{\sma{k_y,\kpar}}.
\end{align}
We now apply that (i) $\bar{\calu}$ is unitary, (ii) exp$(i\noc k_z c/2)$ and $\bar{\calu}$ are analytic functions of $\bk$. To show that $\bar{\calu}$ is analytic, first consider the analyticity of the Bloch Hamiltonian $H(\bk)$ (which is apparent from inspection of Eq.\ (\ref{basisvec}) and (\ref{tightbindingbloch})) and consequently of the occupied-band projection $P(\bk)$, by assumption of a finite gap for all $\bk$. That $\bar{\calu}_{\sma{k_y +2\pi/a\leftarrow k_y}}$ is analytic follows from it being a matrix representation of exp$(-ik_zc/2)\,P(k_y+2\pi/a,\kpar)\,V({-}\tilde{\boldsymbol{b}}_2)\,\ubmx\,P(k_y,\kpar)$, as we defined in Eq.\ (\ref{hooboy}). \\

Together, (i) and (ii) imply that exp$(i\noc k_zc/2)$det[$\bar{\calu}$] is an analytic phase factor, and therefore its phase cannot wind around any contractible loop in the plane; on the other hand, the phase of $\zeta$ will wind around its zeroes, but Eq.\ (\ref{pfaffianresult}) implies that zeros of $\zeta$ always appear as glide-related pairs. Further applying the analyticity of exp$(i\noc k_zc/2)$det[$\bar{\calu}$], we conclude that glide-related zeroes have the same vorticity, which we define by the phase-winding of $\zeta$ around each zero. Each glide-related pair belongs to the same half-glide plane; recall here that the two half planes are defined by $k_z \in [0,\pi/c]$ and $k_z \in [-\pi/c,0]$. Due to time-reversal symmetry, every glide-related pair in one half plane has a partner pair in the other half plane with opposite vorticity, as we illustrate in Fig. \ref{fig:pfaffianzeros}(b). The same figure demonstrates this minimal set of zeros can always mutually annihilate.

\section{Searching for rotationally-inverted topological insulators} \label{rotationalinversion}

To efficiently diagnose topological materials, we propose to search for inversions of the rotational quantum numbers. Such a criterion to diagnose the quantum anomalous Hall effect (QAHE)  is already known\cite{bulktopinvChen} for symmorphic space groups, and in App.\ \ref{app:screw} we generalize this criterion to describe any space group. In App.\ \ref{app:mcnrotate}, we describe how rotational inversion can lead to a nontrivial mirror Chern number, as we exemplify with the KHg$X$ material class.   

\subsection{Quantum anomalous Hall effect due to rotational inversion} \label{app:screw}

Let us consider a space group with an $\bar{n}$-fold rotational symmetry ($C_{\sma{\bar{n},\bdelta}}$); our discussion applies to both screw ($\bdelta \neq 0$) and normal ($\bdelta = 0$) rotations, as well as to integer- and half-integer-spin representations. If nonzero, a spatial origin can always be found where $\bdelta$ lies parallel to the rotational axis,\cite{Lax} which we align in $\vec{z}$. Independent of the spatial origin, the $\vec{z}$-component ($\bdelta_{\shortparallel}$) of $\bdelta$ always satisfies that $\bar{n}\bdelta_{\shortparallel}$ is a lattice vector. This follows first from
\begin{align}
[C_{\sma{\bar{n},\bdelta}}]^{\bar{n}} = \bar{E}\;t\big(\,[I+D_{\bar{n}}+D_{\bar{n}}^2+\ldots+D_{\bar{n}}^{\bar{n}-1}]\bdelta\,\big)
\end{align}
where $\bar{E}$ is a $2\pi$ rotation, $t(\br)$ is a translation by the vector $\br$, and $D_{\bar{n}}$ is the vector representation of an $\bar{n}$-fold rotation in $\R^3$. Noting further that 
\begin{align}
\calk_{\shortparallel} = \frac{I+D_{\bar{n}}+D_{\bar{n}}^2+\ldots+D_{\bar{n}}^{\bar{n}-1}}{\bar{n}}
\end{align}
projects to the rotational axis as $\calk_{\shortparallel}\vec{z}{=}\vec{z}$ and $\calk_{\shortparallel}\vec{x}{=}\calk_{\shortparallel}\vec{y}{=}0$, we obtain
\begin{align} \label{inmoredetail}
[C_{\sma{\bar{n},\bdelta}}]^{\bar{n}} = \bar{E}\,t\big(\,\bar{n}\bdelta_{\shortparallel}\,\big).
\end{align}
Then by applying the closure property of any space group, we conclude that $\bar{n}\bdelta_{\shortparallel}$ must be a lattice vector.\\

 Our goal is to determine the Chern number ($\calc$) in a two-torus normal to $\vec{z}$. For a 2D crystal, this two-torus ($\calt^2$) would be its Brillouin zone (BZ), while for a 3D crystal, the two-torus would be a planar submanifold of the BZ at fixed $k_z$, so that the rotational axis is normal to this plane. We find for a crystal with $\bar{n}$-fold rotational symmetry that the Chern number is determined modulo $\bar{n}$, by the rotational eigenvalues at various high-symmetry momenta. To define these rotational eigenvalues, it will be useful to recall certain notations from App.\ \ref{sec:spacetimetight}: we denote the representation of $\cndel$ in the Bloch-wave orbital basis:
\begin{align}
\hatcndel(\bk) = e^{-i (D_n \boldsymbol{k}) \cdot \bdelta }\, \ucndel, 
\end{align}
as well as in the occupied-band basis:
\begin{align} \label{recallag}
[\,\brevecndel(D_n\bk+\bG,\bk)\,]_{ij} = \bra{u_{i,D_n\bk+\bG}} \,V(-\bG)\, \hatcndel(\bk) \,\ket{u_{j,\bk}}. 
\end{align}
A $C_n$-invariant momentum is defined by $\bar{\bk}=D_n\bar{\bk}$ up to some reciprocal vector ($\bG(\bar{\bk};D_n)$) that depends on $\bar{\bk}$ and $D_n$; the various $\bar{\bk}$ are illustrated in Fig.\ \ref{fig:rotational}. For a crystal whose space group includes a $\cbarndel$ symmetry, there would exist $C_n$-invariant momenta in the corresponding Brillouin zone for any $n$ that divides $\bar{n}$, e.g., for $C_{\sma{4,\bdelta}}$-symmetric crystals, there exist two $C_4$-invariant momenta ($\Gamma$ and $M$) as well as two $C_2$-invariant momenta ($X$ and $Y$), as illustrated in Fig.\ \ref{fig:rotational}(b). Henceforth, we use $\bar{n}$ to label the space-group symmetry of the real-space crystal, and $\{n|(\bar{n}/n){\in}\Z^{+}\}$ to label the little-group symmetries of individual momenta. At each rotationally-invariant momentum, bands may be labelled by quantum numbers $\{\lambda_{\sma{n,\bdelta,i}}(\bar{\bk}) | i=1,\ldots,\noc\}$, which are the eigenvalues of the matrix $\brevecndel(\bar{\bk},\bar{\bk})$. Now we are ready to state our results, with reference to Fig.\ \ref{fig:rotational}: for space groups with
\begin{align} \label{rotationresults}
\text{(i)}\;\; \csixdel \;\;\text{symmetry,} \myspace \myspace & e^{-i\pi \calc/3} = e^{i\noc (6\bk \cdot \bdelta + F\pi)}\;\prod_{i=1}^{\noc} \,\lambda_{\sma{6,\bdelta,i}}(\Gamma)\,\lambda_{\sma{3,2\bdelta,i}}(K)\,\lambda_{\sma{2,3\bdelta,i}}(M), \notag \\
\text{(ii)}\;\; C_{\sma{4,\bdelta}} \;\;\text{symmetry,} \myspace \myspace & e^{-i\pi \calc/2} = e^{i\noc (4\bk \cdot \bdelta + F\pi)}\;\prod_{i=1}^{\noc} \,\lambda_{\sma{4,\bdelta,i}}(\Gamma)\,\lambda_{\sma{2,2\bdelta,i}}(X)\,\lambda_{\sma{4,\bdelta,i}}(M), \notag \\
\text{(iii)}\;\; C_{\sma{3,\bdelta}} \;\;\text{symmetry,} \myspace \myspace & e^{-i2\pi \calc/3} = e^{i\noc (3\bk \cdot \bdelta + F\pi)}\;\prod_{i=1}^{\noc} \,\lambda_{\sma{3,\bdelta,i}}(\Gamma)\,\lambda_{\sma{3,\bdelta,i}}(K_1)\,\lambda_{\sma{3,\bdelta,i}}(K_2), \notag \\
\text{(iv)}\;\; C_{\sma{2,\bdelta}} \;\;\text{symmetry,} \myspace \myspace & e^{-i\pi \calc} = e^{i4\noc \bk \cdot \bdelta}\;\prod_{i=1}^{\noc} \,\lambda_{\sma{2,\bdelta,i}}(\Gamma)\,\lambda_{\sma{2,\bdelta,i}}(X)\,\lambda_{\sma{2,\bdelta,i}}(M)\,\lambda_{\sma{2,\bdelta,i}}(Y).
\end{align}
Here, $F=0$ ($1$) applies to integer-spin (resp.\ half-integer-spin) representations of the space group.\\

\begin{figure}[ht]
\centering
\includegraphics[width=9 cm]{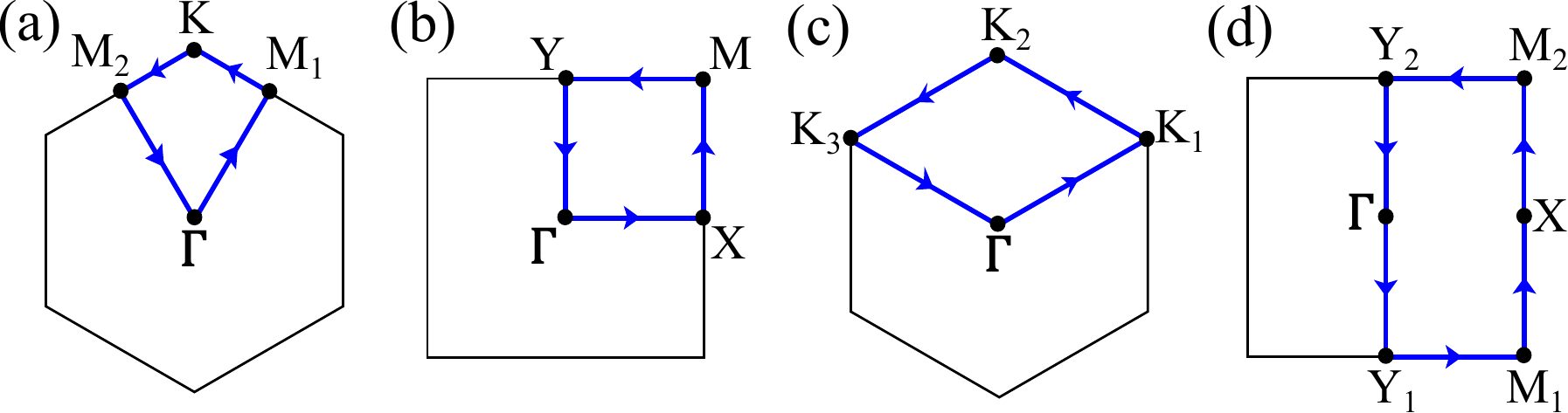}
    \caption{Rotationally-symmetric primitive cells with high-symmetry momenta indicated. Each cell in (a-d) has $C_{\sma{\bar{n},\bdelta}}$ symmetry, with $\bar{n}=6,4,3$ and $2$ respectively. For each cell, a symmetrically-chosen Wilson loop is highlighted in blue, such that it encloses $1/\bar{n}$ of the cell.} \label{fig:rotational}
\end{figure}

It is worth commenting that our rotational-inversion formulae in Eq.\ (\ref{rotationresults}) produces the \emph{absolute} Chern number (mod $\bar{n}$) from knowledge of \emph{all} occupied bands.  It is often easier to compute a \emph{change} in Chern number (mod $\bar{n}$) from knowledge of the inverting bands near the Fermi level, as we illustrate for our material class in App.\ \ref{app:mcnrotate}.  \\

Before we delve into the derivations, it pays to particularize the identity (\ref{uv}) for a reciprocal vector ($\bG$) that is orthogonal to $\vec{z}$:
\begin{align} 
\hatcndel(\bk)\, V(\bG) = V(D_n\bG) \,\hatcndel(\bk);
\end{align}
in particular, this identity holds for any of $\{\bG(\bar{\bk};D_n)\}$. Lastly, we define a Wilson line on this two-torus ($\calt^2$) as 
\begin{align}
[\Wline{\bktwo}{\bkone}]_{mn} = \bra{u_{m,\bktwo}}\,\prod_{\bk}^{\bktwo \leftarrow \bkone} P(\bk)\,\ket{u_{n,\bkone}}, 
\end{align}
where $\bktwo {\leftarrow} \bkone$ indicates that the product of projections is path-ordered from initial point $\bkone$, and the path is chosen in this Appendix as the shortest (in inverse-length units) that interpolates between $\bkone$ and $\bktwo$, as illustrated in Fig.\ \ref{fig:rotational}. From Eq.\ (\ref{symmonhk}) and our definition of $\brevecndel$ in Eq.\ (\ref{recallag}), we further deduce that\cite{berryphaseTCI}
\begin{align} \label{ultimateeq}
\brevecndel(D_n\bktwo,\bktwo)\,\Wline{\bktwo}{\bkone}\,\brevecndel^{\mo}(\bkone,D_n\bkone)   = \Wline{D_n\bktwo}{D_n\bkone}.
\end{align}

\subsubsection{QAHE criterion for systems with six-fold rotational symmetry}
  
The six-fold symmetry constrains the Berry curvature as $\calf(\bk) = \calf(D_6\bk)$, and therefore the Berry flux enclosed by the Wilson loop of Fig.\ \ref{fig:rotational}(a) is $1/6$ of the total flux, i.e., $2\pi \calc$. Then by Stoke's theorem,
\begin{align} \label{Chern6}
e^{-i\pi \calc/3} = \text{det}\big[\;\Wline{\Gamma}{M_2}\,\Wline{M_2}{K}\,\Wline{K}{M_1}\,\Wline{M_1}{\Gamma}\;].
\end{align}
Into this equation, we insert two identities which particularize Eq.\ (\ref{ultimateeq}):
\begin{align}
 \brevecsixdel(\Gamma,\Gamma)\,\Wline{\Gamma}{M_1}\, \brevecsixdel^{\mo}(M_1,M_2) = \Wline{\Gamma}{M_2}, \ins{and} \brevecthreetwodel^{\mo}(M_2,M_1)\,\Wline{M_1}{K}\, \brevecthreetwodel(K,K) = \Wline{M_2}{K}.
\end{align}
Then applying $\Wline{\bk'}{\bk}\Wline{\bk}{\bk'}{=}\Wline{\bk'}{\bk}[\Wline{\bk'}{\bk}]^{\sma{-1}}{=}I$ for any $\bk,\bk'$, we obtain:
\begin{align}
e^{-i\pi \calc/3} = \text{det}\big[\;\brevecsixdel(\Gamma,\Gamma)\,\brevectwothreedel^{\mo}(M_1,M_1)\,\brevecthreetwodel(K,K)\;].
\end{align}
This expression is simplified by relating $\brevectwothreedel^{\mo}(M_1,M_1)$ to $\brevectwothreedel(M_1,M_1)$: since $\ctwothreedel^2$ is a $2\pi$ rotation combined with a $6\bdelta$ translation (if $\bdelta \neq 0$),
\begin{align}
\brevectwothreedel^{\mo}(M_1,M_1) = (-1)^Fe^{i6\bk \cdot \bdelta}\,\brevectwothreedel(M_1,M_1), 
\end{align}
with $F= 0 (1)$ for integer-spin (half-integer spin) representations, and therefore
\begin{align}
e^{-i\pi \calc/3} = e^{i\noc (6\bk \cdot \bdelta + F\pi)}\;\text{det}\big[\;\brevecsixdel(\Gamma,\Gamma)\,\brevectwothreedel(M_1,M_1)\,\brevecthreetwodel(K,K)\;],
\end{align}
which immediately leads to the first equation of (\ref{rotationresults}).

\subsubsection{QAHE criterion for systems with four-fold rotational symmetry}

The other proofs are similar in structure, so we shall be brief. The four-fold symmetry constrains the Berry curvature as $\calf(\bk) = \calf(D_4\bk)$, and therefore
\begin{align}
e^{-i \pi \calc/2} =\det[\;\Wline{\Gamma}{Y}\,\Wline{Y}{M}\,\Wline{M}{X}\,\Wline{X}{\Gamma}\;],
\end{align}
with the Wilson loop drawn in Fig.\ \ref{fig:rotational}(b). Into this equation, we insert two identities:
\begin{align}
\brevecfourdel(\Gamma,\Gamma)\, \Wline{\Gamma}{X}\,\brevecfourdel^{\mo}(X,Y) = \Wline{\Gamma}{Y} \ins{and} 
\brevecfourdel^{\mo}(Y,X)\,\Wline{X}{M}\,\brevecfourdel(M,M) = \Wline{Y}{M},
\end{align}
to obtain
\begin{align}
e^{-i \pi \calc/2} = \det[\; \brevecfourdel(\Gamma,\Gamma)\, \brevectwotwodel^{-1}(X,X)\,\brevecfourdel(M,M)\,\;]
\end{align}
Applying that $C_{\sma{2,2\bdelta}}^2$ is a $2\pi$ rotation combined with a $4\bdelta$ translation (if $\bdelta \neq 0$),
\begin{align}
\brevectwotwodel^{-1}(X,X)=(-1)^F\,e^{i4\bk \cdot \bdelta}\,\brevectwotwodel(X,X),
\end{align}
and we arrive at the second equation of (\ref{rotationresults}).

\subsubsection{QAHE criterion for systems with three-fold rotational symmetry}

The three-fold symmetry gives us that
\begin{align}
e^{-i 2\pi \calc/3} =\det[\;\Wline{\Gamma}{K_3}\,\Wline{K_3}{K_2}\,\Wline{K_2}{K_1}\,\Wline{K_1}{\Gamma}\;],
\end{align}
with the Wilson loop drawn in Fig.\ \ref{fig:rotational}(c). Into this equation, we insert two identities:
\begin{align}
\brevecthreedel(\Gamma,\Gamma)\, \Wline{\Gamma}{K_1}\,\brevecthreedel^{\mo}(K_1,K_3) = \Wline{\Gamma}{K_3} \ins{and} 
\brevecthreedel^{\mo}(K_3,K_1)\,\Wline{K_1}{K_2}\,\brevecthreedel(K_2,K_2) = \Wline{K_3}{K_2},
\end{align}
to obtain
\begin{align}
e^{-i 2\pi \calc/3} = \det[\; \brevecthreedel(\Gamma,\Gamma)\, \brevecthreedel^{-2}(K_1,K_1)\,\brevecthreedel(K_2,K_2)\,\;]
\end{align}
Applying that $\cthreedel^3$ is a $2\pi$ rotation combined with a $3\bdelta$ translation (if $\bdelta \neq 0$),
\begin{align}
\brevecthreedel^{-2}(K_1,K_1)=(-1)^F\,e^{i3\bk \cdot \bdelta}\,\brevecthreedel(K_1,K_1),
\end{align}
and we arrive at the third equation of (\ref{rotationresults}).

\subsubsection{QAHE criterion for systems with two-fold rotational symmetry}

The two-fold symmetry gives us that
\begin{align}
e^{-i \pi \calc} =\det[\;\Wline{\Gamma}{Y_2}\, \Wline{Y_2}{M_2} \,\Wline{M_2}{X}\,\Wline{X}{M_1}\,\Wline{M_1}{Y_1}\,\Wline{Y_1}{\Gamma}\;],
\end{align}
with the Wilson loop drawn in Fig.\ \ref{fig:rotational}(d). Into this equation, we insert the identities:
\begin{align}
\brevectwodel(\Gamma,\Gamma)\, \Wline{\Gamma}{Y_1}\,\brevectwodel^{\mo}(Y_1,Y_2) = \Wline{\Gamma}{Y_2} \ins{and} 
\brevectwodel(M_2,M_1)\,\Wline{M_1}{X}\,\brevectwodel^{\mo}(X,X) = \Wline{M_2}{X},
\end{align}
to obtain
\begin{align} \label{easecomp}
e^{-i \pi \calc} =\det[\;\brevectwodel(\Gamma,\Gamma)\,\brevectwodel^{\mo}(Y_1,Y_2)\, \Wline{Y_2}{M_2} \,\brevectwodel(M_2,M_1)\,\brevectwodel^{\mo}(X,X)\,\Wline{M_1}{Y_1}\;].
\end{align}
It is worth noting that these matrix representations (e.g., $\brevectwodel, \Wline{\bk'}{\bk}$) depend on a particular decomposition of the occupied subspace into $\{u_{\sma{j,\bk}}|j=1,\ldots,\noc\}$, but the final result is independent of this basis choice. Computation of Eq.\ (\ref{easecomp}) is eased if we now choose $|u_{\sma{m,M_1}}\rangle= V(2\pi \vec{y})\,|u_{\sma{m,M_2}}\rangle$ and $|u_{\sma{m,Y_1}}\rangle= V(2\pi \vec{y})\,|u_{\sma{m,Y_2}}\rangle$, such that
\begin{align}
\Wline{M_1}{Y_1} = \Wline{M_2}{Y_2},\myspace \brevectwodel^{\mo}(Y_1,Y_2) = \brevectwodel^{\mo}(Y_2,Y_2) \ins{and} \brevectwodel(M_2,M_1)=\brevectwodel(M_1,M_1).
\end{align}
Further applying that $C_{\sma{2,\bdelta}}^2$ is a $2\pi$ rotation combined with a $2\bdelta$ translation (if $\bdelta \neq 0$),
\begin{align}
\brevectwodel^{-1}(\bar{\bk},\bar{\bk})=(-1)^F\,e^{i2\bk \cdot \bdelta}\,\brevectwodel(\bar{\bk},\bar{\bk}),
\end{align}
and we finally arrive at the last equation of (\ref{rotationresults}).

\subsection{Nontrivial mirror Chern number due to rotational inversion} \label{app:mcnrotate}

Even where there is no net QAHE in the full occupied space, it is still possible to have a QAHE in one mirror subspace.\cite{teo2008} Here, we focus on glideless reflections, since the mirror Chern number is ill-defined for glide reflections. Our strategy is to identify space groups which allow simultaneous eigenstates of rotations and glideless reflections -- only for these space groups may we apply our rotational-inversion formulae (Eq.\ (\ref{rotationresults})) to diagnose this mirror Chern number. It helps to distinguish between normal and screw rotations:\\

\noi{i} Normal rotations ($C_{\sma{\bar{n}}}$) and normal reflections only commute if the rotational and reflection axes coincide; in our convention, both axes would be parallel to $\vec{z}$, and we denote such a reflection by $M_z: z {\rightarrow}-z$. To explain why $[C_{\sma{\bar{n}}},M_{j}]{=}0$ for $j{=}z$ but not $x$ or $y$, it suffices to express $M_{\sma{j}}{=}\cali C_{\sma{2}}$ as a product of a spatial inversion with a two-fold rotation about $\vec{j}$, and applying that inversions commute with any rotation, but two rotations only commute if their axes coincide. We are interested in $M_z$-invariant planes, where each wavevector in said planes is mapped to itself under $M_z$. By assuming that  $C_{\sma{\bar{n}}}$ belongs in the space group, there would exist $C_n$-invariant momenta in these mirror planes such that $n$ divides $\bar{n}$. At each of these $C_n$-invariant momenta, bands may simultaneously carry both $M_z$ and $C_n$ quantum numbers, as deducible from $[M_z,C_{\bar{n}}]{=}0$ (proven above) and $C_{\bar{n}}^m{=}C_n$ for some positive integer $m$. Therefore, the $M_z$ Chern number may be determined modulo $\bar{n}$ through  Eq.\ (\ref{rotationresults}) with $\bdelta{=}0$, if we take the product of rotational eigenvalues only within one $M_z$ subspace.   \\

\noi{ii} Screw rotations and normal reflections do not commute. However, if the reflection and rotational axes coincide, they commute modulo a translation, which in certain representations becomes a trivial phase factor. To elaborate, we have that $\cbarndel\,M_z  = t(2\bdelta_{\shortparallel})\,M_z\,\cbarndel$ for $\bdelta_{\shortparallel}$ the component of $\bdelta$ along $\vec{z}$. In a Bloch-wave representation, $t(2\bdelta_{\shortparallel}) =$ exp $(-i2\bk \cdot \bdelta_{\shortparallel})$ = exp$(-i2k_z|\bdelta_{\shortparallel}|)$. Since $\bdelta_{\shortparallel}$ is not a lattice translation, this phase factor is trivially identity only at the $k_z{=}0$ mirror-invariant plane. In more detail, we have shown in Eq.\ (\ref{inmoredetail}) that $\bar{n}\bdelta_{\shortparallel}$ must be a lattice translation along the rotational axis, i.e., in units where the lattice period in $\vec{z}$ is unity, there are $(\bar{n}-1)$ possible values\cite{Lax} of $\bdelta_{\shortparallel}$ satisfying 
\begin{align}
\bar{n}\bdelta_{\shortparallel} = \bar{m}\vec{z}, \ins{with} \bar{m} \in \{1,2,\ldots, \bar{n}-1\}.
\end{align}
It follows that the phase factor
\begin{align} \label{phasefactor}
e^{-i2\bk \cdot \bdelta_{\shortparallel}}= e^{-i2k_z\bar{m}/\bar{n}}
\end{align}
is unity where $k_z{=}0$, and therefore $\cbarndel$ and $M_z$ representatively commute in this $k_z{=}0$ mirror plane. In space groups where $2\pi \vec{z}$ is a reciprocal vector, there exists another mirror plane at $k_z{=}\pi$ which might be characterized by a mirror Chern number, but here $\cbarndel$ and $M_z$ do not representatively commute, as seen from substituting $k_z{=}\pi$ into Eq.\ (\ref{phasefactor}).\\

Returning to the $k_z{=}0$ mirror plane, we may then determine the mirror Chern number modulo $\bar{n}$ through Eq.\ (\ref{rotationresults}). A case in point is our material class KHg$X$, which is symmetric under a six-fold screw rotation $C_{\sma{6z,c\vec{z}/2}}$, and also under the normal reflection $\bmz = M_{\sma{z,c\vec{z}/2}}$. It is worth clarifying that this symmetry may either be represented as $\bmz$ (with the spatial origin at the inversion center of Fig.\ \ref{fig:eff}(a)), or as a pure reflection $M_z$ (with the origin displaced by $c\vec{z}/4$ from the inversion center); with either choice of origin, $M_z^2=\bmz^2=\bar{E}$ (a $2\pi$ rotation). Modulo six, the mirror Chern number ($\calc_e$) is determined by the rotational eigenvalues in the $M_z=+i$ subspace, which we list in Tab.\ \ref{rottable}; these eigenvalues are directly obtained from ab-initio calculations of the occupied bands, which we label by their symmetry representations in Fig.\ \ref{fig:replabel}. Given that the product of all eigenvalues in Tab.\ \ref{rottable} is exp$({-}i2\pi/3)$, the first line of Eq.\ (\ref{rotationresults}) informs us that $\calc_e=2 $ mod $6$, which we confirm to be just $2$ in the Wilson-loop calculation of App.\ \ref{app:methodswilson}. \\

\begin{table}[ht]
\centering
 \begin{tabular}{C{1cm}C{2cm}C{2cm}C{2cm}}
  \hline
  No. & $\Gamma(C_6)$ & K($C_3$) & M($C_2$) \\
  \hline
   1  &   $ \Gamma_{10} (-i\omega^*) $   &  $K_6(-1)$  &  $M_3(+i)$  \\
   2  &   $ \Gamma_8 (+i\omega^*)  $   &  $K_5(-\omega)$  &  $M_4(-i)$  \\
   3  &   $ \Gamma_9 (+i\omega)  $   &  $K_4(-\omega^*)$  &  $M_3(+i)$  \\
   4  &   $ \Gamma_7   (-i\omega)$   &  $K_4(-\omega^*)$  &  $M_4(-i)$  \\
   5  &   $ \Gamma_{10}(-i\omega^*)$   &  $K_6(-1)$  &  $M_4(-i)$  \\
   6  &   $ \Gamma_8   (+i\omega^*)$   &  $K_4(-\omega^*)$  &  $M_3(+i)$  \\
  \hline
\end{tabular}
\caption{Rotational analysis of KHgSb. For the six occupied bands in the $M_z=+i$ subspace, we list their representation labels and their rotational eigenvalues (in brackets). Note $\omega=$exp$({i{2\pi}/{3}})$, and the product of all eigenvalues in this table is $\omega^{-4}$. \label{rottable}}
\end{table}

\begin{figure}[ht]
\centering
\includegraphics[width=11 cm]{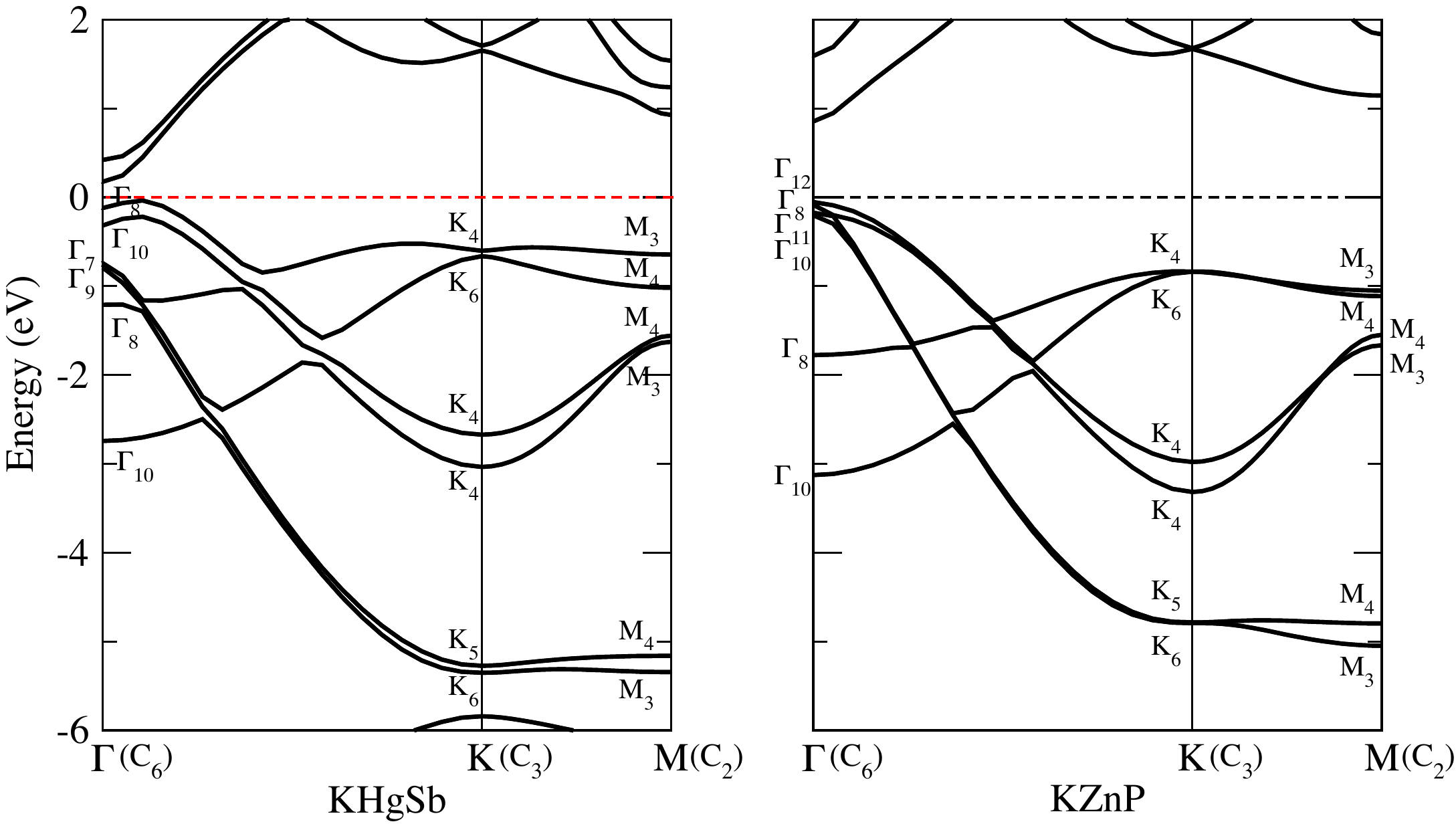}
    \caption{Representation labels~\cite{irrep7,irrep8} of the occupied bands at the high-symmetry points $\Gamma$, $K$ and $M$, for KHgSb (left) and KZnP (right).}\label{fig:replabel}
\end{figure}

While our method requires knowledge of all occupied bands, a shortcut to diagnosis is possible if one has a reference material that one knows to be trivial. For the sake of argument, let us assume we know KZnP to be trivial; its rotational eigenvalues are listed in Tab.\ \ref{rottableKZnP} for direct confirmation. The difference between these two materials lies in a band inversion at $\Gamma$, where s-type $\Gamma_7$ and $\Gamma_9$ orbitals (found in KHgSb) interchange with p-type $\Gamma_{11}$ and $\Gamma_{12}$ (in KZnP). We would like to show that the rotational eigenvalues of these four bands alone determine the change in $\calc_e$ (mod $6$) as a result of the band inversion. It is useful to define the discrete angular momentum ($J_z$) modulo six through $\lambda_{\sma{6,c\vec{z}/2}} \equiv $ exp$(-i\pi J_z/3)$; Tab.\ \ref{rottable} and \ref{rottableKZnP}  inform us that $\Gamma_7$ transform as $J_z=-1/2$, $\Gamma_9$ as $+5/2$, $\Gamma_{11}$ as $+3/2$ and $\Gamma_{12}$ as $-3/2$ -- the net change in angular momentum is $\Delta J_z=2$.  Given that Eq.\ (\ref{rotationresults}) applies individually to KZnP and KHgSb, we divide one equation by the other to obtain $\Delta \calc_e = \Delta J_z =2$ mod $6$.   \\

\begin{table}[H]
\centering
\begin{tabular}{C{1cm}C{2cm}C{2cm}C{2cm}}
  \hline
  No. & $\Gamma(C_6)$ & K($C_3$) & M($C_2$) \\
  \hline
   1  &   $ \Gamma_{10} (-i\omega^*) $   &  $K_6(-1)$  &  $M_3(+i)$  \\
   2  &   $ \Gamma_8 (+i\omega^*)  $   &  $K_5(-\omega)$  &  $M_4(-i)$  \\
   3  &   $ \Gamma_{10}(-i\omega^*)  $   &  $K_4(-\omega^*)$  &  $M_3(+i)$  \\
   4  &   $  \Gamma_{11}(-i)$  &  $K_4(-\omega^*)$  &  $M_4(-i)$  \\
   5  &   $  \Gamma_{8}(+i\omega^*) $   &  $K_6(-1)$  &  $M_4(-i)$  \\
   6  &   $  \Gamma_{12}(+i) $ &  $K_4(-\omega^*)$  &  $M_3(+i)$  \\
  \hline                                  
\end{tabular}
\caption{Rotational analysis of KZnP. For the six occupied bands in the $M_z=+i$ subspace, we list their representation labels~\cite{irrep7,irrep8} and their rotational eigenvalues (in brackets). Note $\omega=$exp$({i{2\pi}/{3}})$, and the product of all eigenvalues in this table is $1$. \label{rottableKZnP}}
\end{table}

\end{appendix}
\end{widetext}

\end{document}